\documentclass[%
reprint,
superscriptaddress,
amsmath,amssymb,amsthm,mathrsfs,
aps,
prl,
]{revtex4-2}

\usepackage{graphicx}
\usepackage{dcolumn}
\usepackage{bm}
\usepackage[usenames]{color}
\usepackage[ 
margin=0.75in,
]{geometry}
\usepackage[normalem]{ulem}

\graphicspath{{./figs/}}
\def\TN{$T_\mathrm{N}$}
\def\Tf{$T_f$}

\def\muSR{$\mu$SR}
\def\muB{$\mu_{\mathrm{B}}$}

\def\znmnte{$\text{Zn}_{0.5}\text{Mn}_{0.5}\text{Te}$}

\allowdisplaybreaks[1]

\begin{document}
	
	\title{Cluster spin glass correlations and dynamics in Zn$_{0.5}$Mn$_{0.5}$Te}
	
	\author{Sabrina R. Hatt}
	\affiliation{ %
		Department of Physics and Astronomy, Brigham Young University, Provo, Utah 84602, USA.
	} %

	\author{Camille Shaw}
	\affiliation{ %
		Department of Physics and Astronomy, Brigham Young University, Provo, Utah 84602, USA.
	} %
	
	\author{Emma Zappala}
	\affiliation{ %
		Department of Physics and Astronomy, Brigham Young University, Provo, Utah 84602, USA.
	} %
	
	\author{Raju Baral}
	\affiliation{ %
		Department of Physics and Astronomy, Brigham Young University, Provo, Utah 84602, USA.
	} %
	\affiliation{ %
		Neutron Scattering Division, Oak Ridge National Laboratory, Oak Ridge, Tennessee 37831, USA.
	} %
	
	\author{Stuart Calder}
	\affiliation{ %
		Neutron Scattering Division, Oak Ridge National Laboratory, Oak Ridge, Tennessee 37831, USA.
	} %
	
	\author{Gerald D. Morris}
	\affiliation{Centre for Molecular and Materials Science, TRIUMF, Vancouver, British Columbia, Canada V6T 2A3}
	
	\author{Brenden R. Ortiz}
	\affiliation{
		Materials Science and Technology Division, Oak Ridge National Laboratory, Oak Ridge, Tennessee 37831, USA.
	}
	
	\author{Karine Chesnel}
	\affiliation{ %
		Department of Physics and Astronomy, Brigham Young University, Provo, Utah 84602, USA.
	} %
	
	\author{Benjamin A. Frandsen}
	\affiliation{ %
		Department of Physics and Astronomy, Brigham Young University, Provo, Utah 84602, USA.
	} %
	\email{benfrandsen@byu.edu}

	\begin{abstract}
		We present a combined magnetometry, muon spin relaxation ($\mu$SR), and neutron scattering study of the insulating spin glass Zn$_{0.5}$Mn$_{0.5}$Te, for which magnetic Mn$^{2+}$ and nonmagnetic Zn$^{2+}$ ions are randomly distributed on a face-centered cubic lattice. The magnetometry and $\mu$SR results confirm a spin freezing transition around $T_f \approx 23$~K, with the spin fluctuation rate decreasing gradually and somewhat inhomogeneously through the sample volume as the temperature decreases toward $T_f$. Characteristic spin correlation times well above $T_f$ are on the order of 10$^{-10}$~s, much slower than typically observed in canonical spin glasses but in line with expectations for a cluster spin glass. Using magnetic pair distribution function (mPDF) analysis and reverse Monte Carlo (RMC) modeling of the magnetic diffuse neutron scattering data, we show that the spin-glass ground state consists of clusters of spins exhibiting short-range-ordered type-III antiferromagnetic correlations, with a locally ordered moment of 3.1(1)~$\mu_{\mathrm{B}}$ between nearest-neighbor spins. The type-III correlations decay exponentially as a function of spin separation distance with a correlation length of approximately 5~\AA. The diffuse magnetic scattering and corresponding mPDF show no significant changes across $T_f$, indicating that the dynamically fluctuating short-range spin correlations in the paramagnetic state retain the same basic type-III configuration that characterizes the spin-glass state; the only change apparent from the neutron scattering data is a gradual reduction of the correlation length and locally ordered moment with increasing temperature. Taken together, these results paint a unique and detailed picture of the local magnetic structure and dynamics in \znmnte\ and provide strong evidence that this material is best described as a cluster spin glass. In addition, this work showcases a new statistical method for extracting diffuse scattering signals from neutron powder diffraction data, which we developed to facilitate the mPDF and RMC analysis of the neutron data. This method has the potential to be broadly useful for neutron powder diffraction experiments on a variety of materials with short-range atomic or magnetic order.

	\end{abstract}
	
	\maketitle
	
	\section{Introduction}	
	A spin glass is a magnetic state in which the magnetic moments are frozen in a disordered configuration, lacking the long-range order typical of conventional magnetic ground states. Such a state can arise in materials that possess magnetic frustration and an element of randomness, e.g. disordered atomic site occupation in an alloy~\cite{binde;rmp86}. Upon warming through the spin-glass freezing temperature \Tf, the system undergoes a classical phase transition into a paramagnetic state with fluctuating magnetic moments~\cite{mydos;rpp15}. In canonical spin glasses such as \textit{Cu}Mn and\textit{Au}Fe, the basic units undergoing the cooperative freezing transition are individual atomic spins; in contrast, cluster spin glasses have as their basic units finite-sized clusters of spins that are locally correlated within a given cluster~\cite{mydos;rpp15}. Despite the lack of long-range magnetic order in both canonical and cluster spin glasses, the magnetic moments nevertheless exhibit short-range correlations as dictated by the magnetic interactions governing the system, and these correlations may persist dynamically above \Tf~\cite{binde;rmp86}. The nature of these correlations and the details of the dynamics above \Tf\ are, however, expected to be distinct in canonical spin glasses compared to cluster spin glasses, with slower dynamics observed in cluster spin glasses~\cite{roych;aipa23}.
	
	Questions about the nature of the spin-glass phase transition have stimulated a vast amount of theoretical and experimental research for over five decades and continue even now~\cite{binde;rmp86, mydos;rpp15, huang;jmmm85, weiss;rmp93, newma;jpcm03}, earning spin glasses a prominent place within condensed matter physics. Indeed, the importance of spin glasses extends well beyond the bounds of traditional condensed matter physics: as a paradigm of complexity, concepts developed in the study of spin glasses have seen application to topics as wide-ranging as protein dynamics, neural networks, combinatorial optimization, and more~\cite{stein2013spin}.
	
	
	
	Experimental probes that are sensitive to local magnetic correlations in the absence of long-range order are crucial for studying spin glasses. Two such probes that have been highly influential in spin-glass studies are muon spin relaxation/rotation (\muSR), in which spin-polarized muons implanted in the sample interact with the local magnetic field they experience, and neutron scattering, which probes the two-spin correlation function in the material. The short-range magnetic correlations in a spin glass produce a diffuse neutron scattering pattern that encodes information about the local magnetic structure.
	
	During the nascent stages of the spin-glass field, crucial advances were enabled by both \muSR\ and neutron scattering~\cite{meneg;pr57, uemur;prb85, binde;rmp86}. In the decades since these seminal early works, technical capabilities for \muSR\ and especially neutron scattering have advanced tremendously, providing data with much greater sensitivity and accuracy than was previously possible~\cite{ehler;i22, parke;acspca24}. For example, increased neutron flux, more widespread use of spin-polarized neutrons, higher energy resolution, and more sophisticated computational modeling capabilities have revolutionized neutron scattering studies of magnetic materials.
	
	The transformative capabilities of modern experimentation and data analysis invite a renewed look into spin glasses, both classic systems from decades past as well as newer systems of interest. A handful of recent neutron-scattering studies of spin glasses highlight the unprecedented level of detail that can be achieved, such as directionally-resolved magnetic correlation lengths in anisotropic systems~\cite{graha;prm20}, precise determination of magnetic cluster sizes~\cite{zhang;prb19}, and subtle differences between static versus dynamic spin correlations across the freezing temperature~\cite{li;prb23, labar;prb21}.
	
\begin{figure}
		\includegraphics[width=75mm]{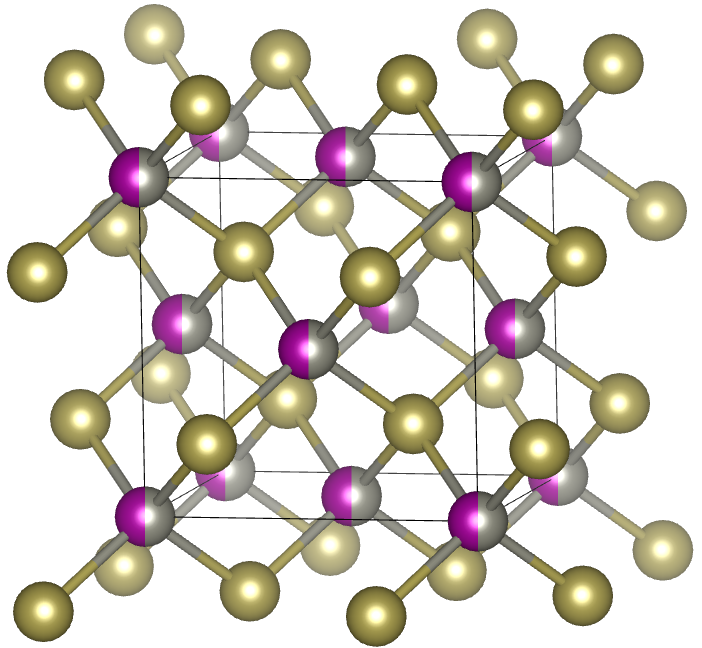}
		\caption {Structure of \znmnte{}, generated in VESTA \cite{momma2011vesta}. The lattice exhibits a zincblende structure composed of two interlaced FCC lattices, one containing only Te atoms (gold) and the other with 50\% each occupancy of Zn and Mn (violet and silver).}
		\label{fig:struc}
	\end{figure}

	
	In this work, we revisit the spin-glass system (Zn,Mn)Te, which is a model system for concentrated, insulating spin glasses with dominant short-range interactions between local magnetic moments~\cite{holde;prb82,mcali;prb84,furdy;jap87,shand;prb98}. For Mn concentrations less than 0.86, (Zn,Mn)Te crystallizes in the zincblende structure (shown in Fig.~\ref{fig:struc}), with the Zn$^{2+}$ and Mn$^{2+}$ ions randomly occupying the positions of a face-centered cubic (fcc) lattice. The freezing temperature for (Zn,Mn)Te can be tuned by adjusting the Zn:Mn ratio; for \znmnte, the composition we investigate in the present work, \Tf\ is around 21--24~K. Earlier neutron diffraction studies revealed a broad distribution of diffuse intensity centered around (1, 1/2, 0) and similar positions in reciprocal space, which was suggested to result from incipient type-III antiferromagnetic ordering~\cite{holde;prb82}. Qualitatively similar diffuse scattering patterns were observed above \Tf, indicating that dynamic short-range magnetic correlations survive in the paramagnetic state with the same overall spatial modulation observed in the frozen state.
	One study on a single crystal specimen~\cite{ono;jpcs99} reported a smaller diffuse neutron scattering signal corresponding to type-I AFM correlations coexisting with the stronger type-III AFM correlations, but this result has not been confirmed by other experiments on single crystals~\cite{giebu;prb89}, leaving the possibility of type-I correlations unsettled. 
	
	The spin-glass behavior of \znmnte\ has been established through observation of the characteristic bifurcation of field-cooled and zero-field-cooled magnetometry data at the transition and the frequency-dependent shift of the maximum in the ac magnetic susceptibility~\cite{mcali;prb84, shand;prb98}. In addition, the magnetic specific heat shows a continuous evolution as a function of temperature across the transition in the closely related compounds (Cd,Mn)Te~\cite{galaz;prb80} and (Zn,Mn)Se~\cite{tward;prb87}, providing further confirmation of a spin-glass transition. Although not widely discussed in the literature on this system, some authors proposed a cluster spin glass scenario for (Cd,Mn)Te and (Zn,Mn)Te~\cite{furdy;jap87} based on the neutron scattering data. It was pointed out, however, that microscopic details about the spin clusters, intra-cluster correlations, and cluster dynamics were scarce~\cite{furdy;jap88}.  
	
	
	Here, we combine magnetometry, \muSR, and magnetic pair distribution function (mPDF)~\cite{frand;aca14, frand;aca15} and reverse Monte Carlo (RMC)~\cite{paddi;prl12, paddi;jpcm13} analysis of neutron scattering data on \znmnte{} to provide a highly detailed view of the evolution of the local magnetic correlations and their dynamics~in \znmnte{} from deep within the spin-glass state up to room temperature. Key results include confirmation of dominant short-range type-III antiferromagnetic ordering, with no evidence for coexisting or competing type-I correlations; quantitative refinement of the locally ordered magnetic moment, spin direction, and magnetic correlation length as a function of temperature; measurement of the homogeneity of the spin-freezing process throughout the sample; demonstration that the instantaneous local magnetic structure undergoes no significant change across \Tf; and observation of slow magnetic dynamics consistent with the scenario of \znmnte\ as a cluster spin glass rather than a canonical spin glass.~In addition, we describe a data processing method that effectively separates the diffuse magnetic scattering signal from the sharp nuclear Bragg peaks in the neutron data, enabling reliable mPDF analysis of data collected from a neutron powder diffractometer not specifically designed for conventional PDF analysis. The results shed new light on this class of spin-glass systems and demonstrate the effectiveness of this methodological approach for studying spin glasses and other systems with short-range magnetic order. 
	
	\section{Methods}
	
	\subsection{Sample Synthesis}
	Stoichiometric amounts of Zn powder (99.9\%), Mn powder (99.95\%), and Te pieces (99.999\%) were mixed thoroughly with an agate mortar and pestle inside an argon glovebox and sealed in an evacuated quartz ampoule. The ampoule was heated to 700~$^{\circ}$C and held there for 24 hours. The sample was then removed from the ampoule, ground in an agate mortar and pestle, sealed once again in an evacuated quartz ampoule, and annealed again at 700~$^{\circ}$C for 24 hours. Four samples of approximately 1~g each were prepared in this way and combined into a single sample to be used for subsequent experimentation.
	
	\subsection{Bulk Structural and Magnetic Characterization}
	
	The sample structure and purity were quantified via laboratory X-ray diffraction (XRD) using a PANalytical X’Pert Pro reflection geometry powder XRD instrument, and the resulting data were analyzed by Rietveld refinement using PANalytical's High Score Plus software \cite{degen;pd14}. The bulk magnetic behavior was characterized with a superconducting quantum interference device (SQUID) magnetometer (Quantum Design MPMS-3). Measurements were made in a warming sequence between 1.8~K and 300~K under zero-field-cooled (ZFC) and field-cooled (FC) conditions with an applied field of 1000~Oe. 
	
	We also gathered X-ray pair distribution function (PDF) data using Brookhaven National Laboratory's PDF beamline 28-ID-1 and modeled the data in PDFgui~\cite{farro;jpcm07} to investigate the local atomic structure. The powder sample was loaded into a polyimide capillary and the total scattering intensity was recorded on a PILATUS3 X 2M CdTe Detector. Data processing, including background subtraction, was performed using standard scripts on the beamline, and the real-space PDF pattern was generated using xPDFsuite~\cite{yang;arxiv15} with a maximum momentum transfer of 25.5~\AA$^{-1}$. 
	
	\subsection{Muon Spin Relaxation}
	
	In a \muSR\ experiment, spin-polarized positive muons with a mean lifetime of about 2.2~$\mu$s are implanted in a sample, and their decay into positrons is monitored as a function of time after implantation. The decay positron is emitted preferentially in the direction of the muon spin at the moment of decay. The experimental quantity of interest is the time-dependent \muSR\ asymmetry, which is the normalized difference in positron counts between two detectors placed on opposite sides of the sample as a function of time after muon implantation. This quantity is proportional to the muon ensemble spin polarization projected along the axis defined by the two detectors~\cite{hilli;nrmp22}. As a probe of the local magnetic environment, \muSR\ is sensitive both to long-range and short-range magnetic correlations and can distinguish between sample volumes that exhibit static magnetism versus paramagnetism. In addition, \muSR\ can probe magnetic relaxation times between approximately 10$^{-5}$ and 10$^{-11}$~s, which is situated nicely between the slower time-scale sensitivity of SQUID magnetometry and the faster sensitivity of neutron diffraction~\cite{hilli;nrmp22}.
	
	\muSR\ experiments were performed at TRIUMF Laboratory using the LAMPF spectrometer on endstation M20D. A 500-mg portion of the powder sample used for the neutron diffraction experiment was pressed into a pellet of diameter approximately 1~cm for use in the \muSR\ experiment. The temperature was controlled using a helium gas flow cryostat with a base temperature of 1.9~K. Helmholtz coils were used to generate a uniform magnetic field at the sample position, with a maximum applied field of 0.4~T. The \muSR\ spectra were analyzed using the open-source software \texttt{musrfit}~\cite{suter;physproc12}.
	
	
	
	
	\subsection{Neutron Diffraction Experiment and Data Analysis}
	
	\subsubsection{Neutron Diffraction Experiment}
	Neutron diffraction measurements were performed on the HB2A powder diffractometer at Oak Ridge National Laboratory's High Flux Isotope Reactor (HFIR) facility. The sample was loaded into a vanadium sample can and mounted in a helium cryostat. Diffraction patterns were collected in a warming sequence at 3~K, 15, 28, 40, 100, 150, 225, and 295~K for 6 hours each. The neutron wavelength used was 1.54 \AA, yielding a maximum momentum transfer $Q$ of approximately 8~\AA$^{-1}$. At 3~K, an additional diffraction pattern was collected with a neutron wavelength of 1.12~\AA$^{-1}$, yielding a maximum $Q$ of about 10~\AA$^{-1}$. An empty sample can was measured with both wavelengths for background subtraction. The scattered intensity was normalized to barns per steradian per atom using the scale factor from GSAS-II Rietveld refinements~\cite{paddi;jac25}. As shown in the Supplemental Material~\footnote{See Supplemental Material at [URL will be inserted by publisher] for analysis of previously published ac magnetometry data, Curie-Weiss fits, further verification of the normalization of the neutron diffraction data and the estimated locally ordered magnetic moment, code for the peak removal algorithm, and the data files used in this study.}, the normalization was checked and found to be consistent with comparison to the high-$Q$ scattering limit and the paramagnetic scattering expected for \znmnte.  
	
	
	\subsubsection{Magnetic PDF Analysis}
	
	Our primary approach for extracting information about the local spin correlations from the neutron scattering data is to use the magnetic pair distribution function (mPDF)~\cite{frand;aca14,frand;aca15}, which entails computing the Fourier transform of the magnetic scattering pattern to yield the spin-pair correlation function directly in real space. The successful use of the Fourier transform of diffuse magnetic scattering to study spin glasses has been demonstrated in the past~\cite{wiede;ssc81, wiede;jpcss85, wills;prb01, ehler;prb10}, and is a promising method for this project. In recent years, a complete formalism of mPDF analysis has been developed~\cite{frand;cm24}, enabling a more quantitative treatment of Fourier-transformed diffuse scattering data. Together with major improvements in neutron instrumentation and data quality, these developments have enabled successful mPDF studies of short-range magnetism in systems ranging from geometrically frustrated quantum magnets~\cite{frand;prm17, lefra;prb19, dun;prb21, qures;prb22} to magnetically enhanced thermoelectrics~\cite{baral;matter22, frand;jap22} to spin glasses~\cite{roth;iucrj18, zhang;prb19}. 
	
	Magnetic PDF data were obtained via Fourier transformation of the $Q$-dependent diffuse magnetic scattering cross section. The procedure to separate the magnetic scattering from the nuclear scattering, newly developed for this study, is described in detail in the Results section. Two variants of the mPDF were utilized in this study. The first, which we call the deconvolved or normalized mPDF, is given by~\cite{frand;aca14, kodam;jpsj17}
	\begin{widetext}
		\begin{align}
			G_{\mathrm{mag}}(r)&=\frac{2}{\pi}\int_{Q_{\mathrm{min}}}^{\infty} Q\left(\frac{\left(\text{d}\sigma/\text{d}\Omega\right)_{\mathrm{mag}}}{\frac{2}{3}N_sS(S+1)(\gamma r_0)^2 [f_m(Q)]^2}-1\right)\sin{(Q r)} \text{d}Q \label{eq:FT}
			\\&=\frac{3}{2 S(S+1)}\left(\frac{1}{N_s}\sum\limits_{i \ne j}\left[ \frac{A_{ij}}{r}\delta (r-r_{ij})+B_{ij}\frac{r}{r_{ij}^3}\Theta (r_{ij}-r)\right] - 4\pi r \rho_0 \frac{2}{3} m^2\right) \label{eq:fullfofr}.
		\end{align}
	\end{widetext}
	Eq.~\ref{eq:FT} provides the experimental definition of the mPDF, while Eq.~\ref{eq:fullfofr} shows how the mPDF is calculated for a given configuration of spins. 
	In these equations, $Q_{\mathrm{min}}$ is the minimum momentum transfer included in the Fourier transform (assumed to exclude the small-angle scattering regime), $\left(\mathrm{d}\sigma/\mathrm{d}\Omega\right)_{\mathrm{mag}}$ is the magnetic differential scattering cross section, $r$ is the distance in real space, $r_0=\frac{\mu _0}{4\pi}\frac{e^2}{m_e}$ is the classical electron radius, $\gamma = 1.913$ is the neutron magnetic moment in units of nuclear magnetons, $S$ is the spin quantum number in units of~$\hbar$, $f_m(Q)$ is the magnetic form factor, $N_s$ is the number of spins in the system, $i$ and $j$ label individual spins~$ \mathbf{S_{\textit i}}$ and~$\mathbf{S_{\textit j}}$ separated by the distance~$r_{ij}$, and $A_{ij}$ and $B_{ij}$ are spin orientation coefficients as described in \cite{frand;aca14}. In addition, $\Theta$ is the Heaviside step function, $m$ is the average magnetic moment in Bohr magnetons (which is zero for anything with no net magnetization, such as antiferromagnets), and $\rho_0$ is the number of spins per unit volume. We refer to it as the deconvolved mPDF because division by $[f_m(Q)]^2$ in Eq.~\ref{eq:FT} effectively deconvolves the real-space mPDF from the effect of the $Q$-space damping of the magnetic intensity caused by the magnetic form factor.
	
	The other mPDF variant used, which we refer to as the non-deconvolved or unnormalized mPDF, is given by~\cite{frand;aca15}
	\begin{align}
		d_{\mathrm{mag}}(r)&=\frac{2}{\pi}\int_{Q_{\mathrm{min}}}^{\infty} Q\left(\frac{\text{d}\sigma}{\text{d}\Omega}\right)_{\mathrm{mag}}\sin{(Q r)} \text{d}Q\label{eq;drexp}
		\\&=C_1 \times G_{\mathrm{mag}}(r)\ast S(r) + C_2 \times \frac{\textrm{d}S}{\textrm{d}r},\label{eq;dr}
	\end{align}
	where $C_1$ and $C_2$ are constants, $\ast$ represents the convolution operation, and $S(r)=\mathcal{F}\left\{f_{m}(Q)\right\}\ast \mathcal{F}\left\{ f_{m}(Q)\right\}$ (where $\mathcal{F}$ denotes the Fourier transform). The second term on the right of Eq.~\ref{eq;dr} comes from the self-scattering contribution to the magnetic differential scattering cross section and results in a peak at low $r$ below approximately 1~\AA. As seen from the equations, the non-deconvolved mPDF $d_{\mathrm{mag}}$ is essentially the deconvolved mPDF $G_{\mathrm{mag}}$ convolved twice with the Fourier transform of the magnetic form factor. As a result, it has reduced resolution in real space, so it is advantageous to use $G_{\mathrm{mag}}$ when the quality of the scattering data is high enough. In practice, division by $[f(Q)]^2$ can introduce significant errors into the Fourier transform for noisy scattering data, in which case the non-deconvolved mPDF may be preferable.
	
	The experimental mPDF patterns were produced by the diffpy.mpdf software~\cite{frand;jac22} using the procedure described in Ref.~\cite{frand;jap22}. Both the deconvolved and non-deconvolved mPDF patterns were produced for the data collected at 3~K; for all other temperatures, only the non-deconvolved mPDF patterns were used. We used $Q_{max}=4.5$~\AA$^{-1}$ for the lowest-temperature data set and reduced $Q_{max}$ linearly with temperature to 3.5~\AA$^{-1}$ for the highest-temperature data set, since the weaker magnetic scattering yielded a smaller usable data range. Fits to the mPDF data were performed using diffpy.mpdf. The fitting parameters included the magnitude of the locally ordered magnetic moment, the angles describing the overall orientation of the magnetic configuration relative to the crystal axes, the finite correlation length governing the extent of the short-range magnetic order, and a scale factor for the self-scattering contribution to the mPDF, which is a peak below $r\sim 1$~\AA.

	\subsubsection{Reverse Monte Carlo Modeling}
	
	We complemented the real-space mPDF analysis with analysis in reciprocal space using the reverse Monte Carlo (RMC) method~\cite{paddi;prl12,paddi;jpcm13} to fit supercell spin configurations to the scattering data as implemented in Spinvert \cite{paddison2013spinvert}. In this approach, the orientations of magnetic moments in a large supercell are randomly adjusted until they produce a simulated scattering pattern that is consistent with the data. Considering that the RMC method utilizes reciprocal-space data and does not require a starting guess for the magnetic structure, this method is complementary to the model-dependent real-space mPDF method. 
	
	From the RMC-generated spin configurations, the spin correlation function can be calculated as
	\begin{equation}
		\langle \mathbf{S} (0) \cdot \mathbf{S} (r) \rangle = \frac{1}{N}\sum\limits_{i=1}^{N}\sum\limits_{j=1}^{Z_i(r)}{\frac{\mathbf{S}_i \cdot \mathbf{S}_j}{\langle Z_i(r) \rangle}},
		\label{eq:spincorrelation}
	\end{equation}
	where each magnetic moment is represented as $\mathbf{S}_i$, $\langle Z_i(r) \rangle$ is the average coordination number for a distance $r$ from the central magnetic moment, and $N$ is the total number of magnetic moments in the supercell. Spinvert can be run multiple times to produce several fits. Then the spin correlation function can be calculated and averaged over all fits to give the best result. This function can be used to help visualize the relative orientation of correlated magnetic moments, thus helping to deduce the structure. 
	
	
	\subsection{Comparison of Time-Scales Probed by Relevant Techniques}
	Because the spin-glass transition is a dynamical phenomenon, it is important to understand the difference in time-scales probed by the three primary techniques we use in this study to investigate the magnetic properties of \znmnte, namely magnetometry, \muSR, and neutron diffraction~\cite{hilli;nrmp22}. For dc magnetometry, no measurable signal is produced by spin fluctuations with characteristic relaxation times significantly shorter than the measurement time, meaning that spin relaxation times shorter than approximately 10$^{-1}$--10$^{-2}$~s will be ``invisible'' to common dc magnetometry mesaurements. In contrast, ac magnetometry can be sensitive to spin fluctuations with relaxation times as short as 10$^{-4}$--10$^{-5}$~s. For \muSR, spin relaxation times longer than about 10$^{-5}$~s appear static, while relaxation times significantly shorter than about 10$^{-11}$~s will be too short to be detected. The neutron experiments discussed in this work produce effectively energy-integrated neutron scattering patterns, since the diffractometer employs no energy analysis of the scattered neutrons. Considering the incident neutron wavelength and energy (1.54~\AA, 34.6~meV), any spin fluctuations with relaxation times longer than approximately 10$^{-13}$~s will appear static, and faster fluctuations will go undetected. As will be discussed subsequently, with the energy range of magnetic fluctuations in \znmnte\ limited to about 18~meV and the spectral weight predominantly at even smaller energies~\cite{giebu;prb89, henni;prb02}, the neutron diffraction measurements are effectively probing the instantaneous spin correlations in \znmnte. Considering their different time-scales, we expect the spin-glass transition and nature of the spin correlations to manifest differently for these various techniques.
	
	\section{Results}
	
	\subsection{XRD, Magnetometry, and X-ray PDF}
\begin{figure}
		\includegraphics[width=8cm]{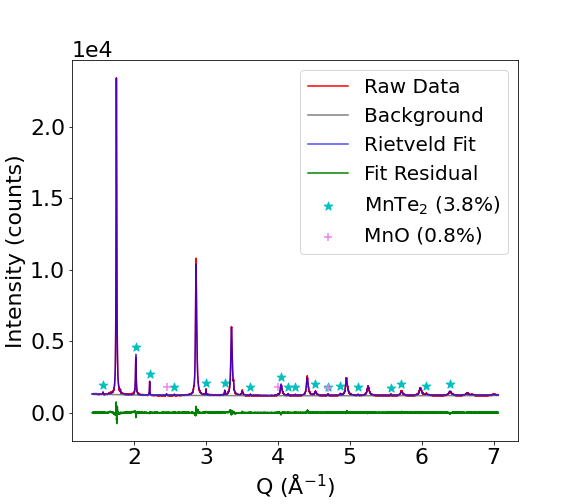}
		\caption {Rietveld refinement of X-ray diffraction (XRD) data. }
		\label{fig:RietveldXRD}
	\end{figure}

	The powder XRD pattern of the sample at room temperature is shown in Fig. \ref{fig:RietveldXRD}.
	The main \znmnte{} phase with the expected cubic structure accounts for 95\% of the molar fraction of the sample. The refined lattice parameter is 6.2150(5)~\AA. Based on an earlier compositional study of the (Zn,Mn)Te system~\cite{furdy;jssc83}, this lattice parameter corresponds to 48\% Mn concentration, which is close to the 50\% composition we targeted in the synthesis. We also identified two minor impurity phases: MnTe$_2$ at 3.8\% and MnO at 1\%. 
	
\begin{figure}
		\includegraphics[width=7.5cm]{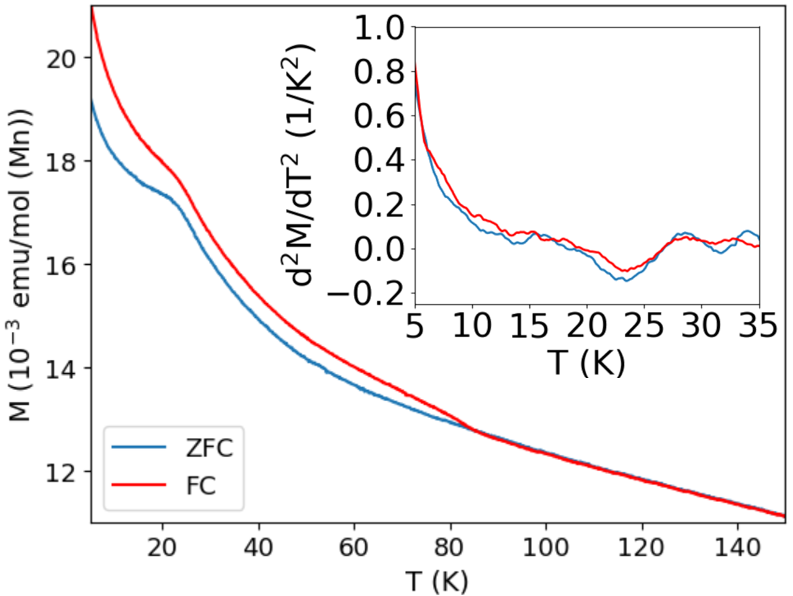}%
		\caption{Field cooled (red) and zero-field cooled (blue) magnetization, gathered by superconducting quantum interference device (SQUID) magnetometry. Data were collected in a warming sequence. Inset: Second derivative of magnetization (normalized to maximum value of 1). We note that percent-level magnetic impurity phases obscure the typical temperature dependence expected for a spin glass, but clear signatures of the spin-glass transition nevertheless remain, as explained in the main text.}%
		\label{fig:Magnetometry}
	\end{figure}

	We display the dc magnetometry data in Fig.~\ref{fig:Magnetometry}.
	The data were collected in a warming sequence. The two most notable features are a kink around 21--24~K and a merging of the ZFC and FC curves around 85~K. The kink agrees closely with the previously reported freezing temperature of 21~K for a very similar composition~\cite{shand;prb98}. The transition is also quite clear in the second derivative of the magnetization (inset of Fig.~\ref{fig:Magnetometry}). Based on the finite width of the kink in the magnetization, we estimate the freezing temperature of our sample to be $T_f = 23(1)$~K. We attribute the ZFC and FC branching around 85~K to the impurity phase MnTe$_2$, which has an antiferromagnetic ordering temperature of 86.5~K~\cite{burle;prb97}. The overall increase of the magnetization with decreasing temperature is more pronounced than previously published data~\cite{shand;prb98}, suggesting that dilute paramagnetic impurities in the sample may contribute to this effect, obscuring somewhat the typical downturn or leveling of the susceptibility below the transition. Fortunately, the XRD and magnetometry both confirm the expected behavior in the majority \znmnte\ phase, affirming the suitability of this sample for detailed studies of the local magnetic correlations using neutron diffraction and \muSR. A Curie-Weiss fit (see Supplemental Material) returns a Curie-Weiss temperature of $\theta_{CW}=-300(2)$~K, indicating strong AFM interactions and a high degree of frustration, with the frustration index $|\theta_{CW}/T_f|\sim13$. The effective moment determined from the Curie-Weiss fit is 6.13(5)~\muB, close to the expected value of $2\sqrt{S(S+1)}=5.92$~\muB\ for $S=5/2$ spins.
	
\begin{figure}
		\includegraphics[width=8cm]{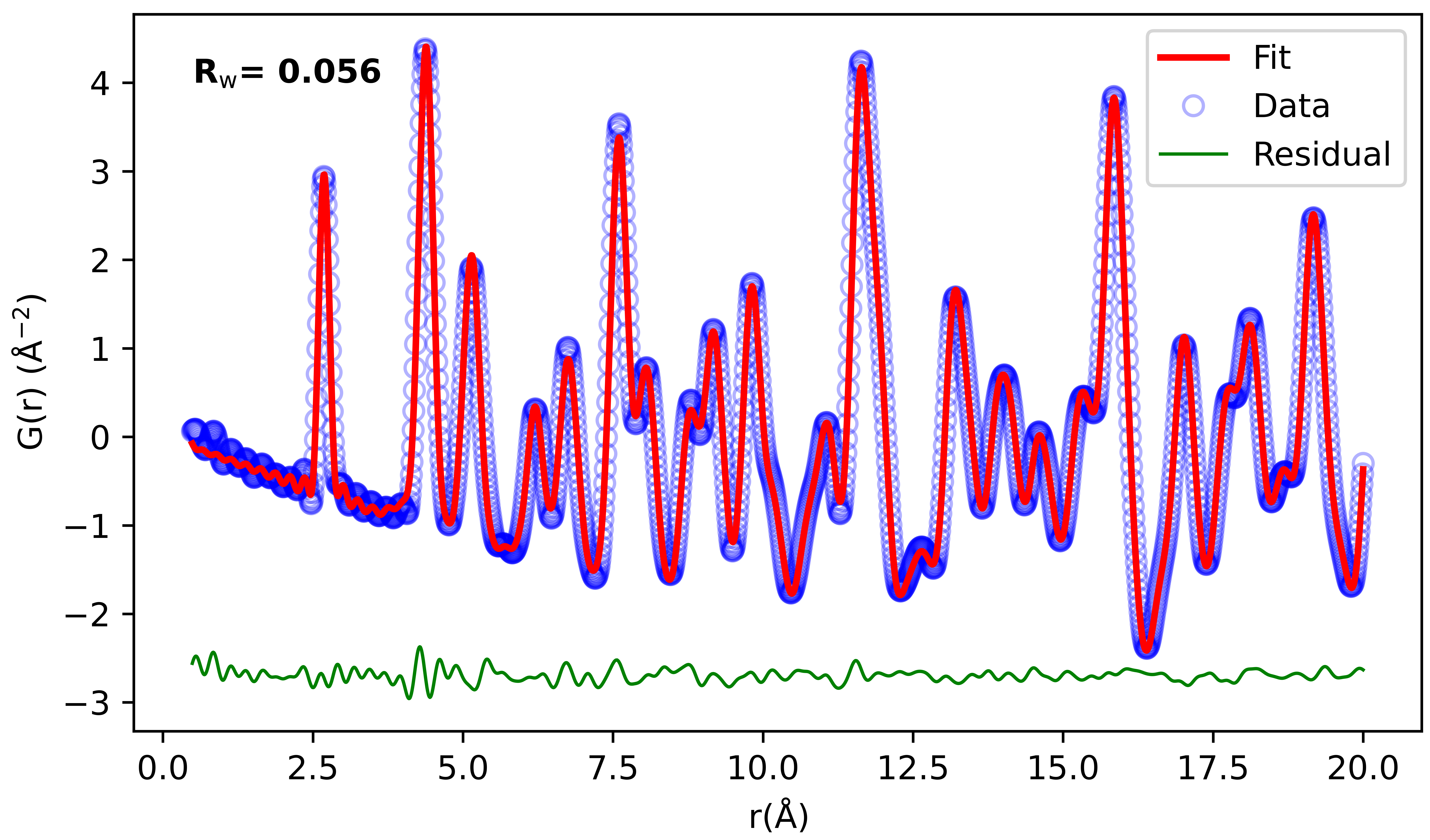}%
		\caption{ X-ray PDF fit and residual at 25~K using a model that assumes a uniform distribution of Zn and Mn atoms, consistent with the average structure. Small features are visible at the interlattice and intralattice nearest and next-nearest neighbor distances, which are determined to not be detrimental to the magnetic structure analysis. }%
		\label{fig:PDF}
	\end{figure}

	PDF analysis provides a useful way to probe the local structure of the sample, which is important considering the role of structural disorder and/or chemical short-range order in many spin glasses. We fit a model of the average zinc-blende structure to the X-ray PDF data, in which each Zn/Mn site is assumed to be occupied by exactly 50\% of a Zn atom and 50\% of a Mn atom, corresponding to a uniform and random distribution of Zn and Mn atoms. We also included minority phases of MnTe$_2$ and MnO in the model. A representative fit to the data collected at 25~K is shown in Fig.~\ref{fig:PDF}.
	The fit quality is excellent, as evidenced by the low value of goodness-of-fit metric $R_w$, which is 0.056 for the fit shown. The molar fraction of \znmnte\ came to 97\%, MnTe$_2$ came to 3\%, and the fit was unable to resolve any MnO content; these values are close to the findings from conventional XRD and the neutron diffraction Rietveld refinements discussed later. The fit residual exhibits a small feature at the (Zn/Mn)-(Zn/Mn) and Te-Te distance around 4.4~\AA\ arising from the calculated peak being slightly narrower and shifted to slightly higher $r$ than the observed peak; this could be a signature of steric effects resulting from the somewhat different ionic radii of Zn$^{2+}$ and Mn$^{2+}$. Regardless, no evidence for significant chemical short-range order is seen in the data. For example, significant clustering of Mn atoms could be expected to distort the local structure toward the hexagonal structure observed in bulk MnTe~\cite{kunit;jdp64}, leading to a noticeable mismatch between the cubic model and data, but no such effect is seen. Neutron PDF would be preferable due to the increased scattering contrast between Mn and Zn, but from the X-ray PDF data available to us, the average-structure model of \znmnte\ assuming a uniform distribution of cations is consistent with observations. We further note that there is very little temperature dependence to the fit quality, with the $R_w$ value varying only between 5\% and 7\% for all data we collected, spanning the temperature range 25 -- 300~K.
	

	\subsection{Muon Spin Relaxation}
	
\begin{figure}
		\includegraphics[width=80mm]{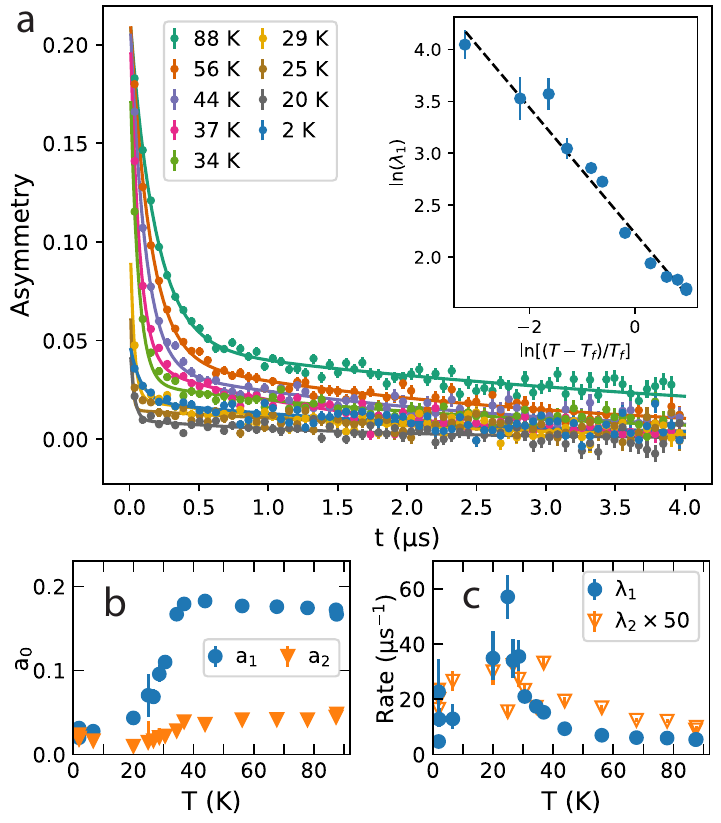}
		\caption{\label{fig:musr} (a) Zero-field \muSR\ asymmetry spectra for \znmnte\ collected between 2~K and 88~K. Colored symbols represent the experimental data, solid curves the fits using Eq.~\ref{eq:muSR}. Inset: Log-log plot of the fast relaxation rate $\lambda_1$ versus the reduced temperature $(T-T_f)/T_f$ above the freezing temperature. (b) Temperature dependence of the fast-relaxing ($a_1$) and slow-relaxing ($a_2$) asymmetry amplitudes. (c) Temperature dependence of the fast relaxation rate $\lambda_1$ and slow relaxation rate $\lambda_2$, where $\lambda_2$ is multiplied by 50 for convenience.}
	\end{figure}

	To characterize the spin-glass transition more fully, we performed \muSR\ measurements of \znmnte\ at several temperatures between 2~K and 88~K. Representative \muSR\ asymmetry spectra collected in zero applied field (ZF) are shown in Fig.~\ref{fig:musr}(a).
	The spectra exhibit two distinct components: a rapidly relaxing front end comprising a majority of the total asymmetry, followed by a long-time tail with much slower relaxation. To quantify this behavior, we performed fits to the spectra using the asymmetry function
	\begin{equation}
		\label{eq:muSR}
		a(t) = a_1 e^{-\lambda_1 t} + a_2 e^{-\lambda_2t},
	\end{equation}
	where $a_1$ and $a_2$ are extrapolated asymmetry values at $t=0$ for each component and $\lambda_1$ and $\lambda_2$ are the fast and slow relaxation rates, respectively. The fits are shown as the solid curves in Fig.~\ref{fig:musr}(a), demonstrating good agreement with the data. We show the best-fit values of the asymmetry parameters and relaxation rates in Fig.~\ref{fig:musr}(b, c). As the temperature decreases below $\sim$34~K, $a_1$ drops sharply, eventually plateauing below 20~K. This behavior of the asymmetry arises from the growth of regions of spins throughout the sample volume that are frozen on the muon timescale, indicating that the freezing transition occurs somewhat inhomogeneously through the sample over this temperature interval.  This may be due in part to local, random fluctuations in the Mn concentration, since the freezing temperature depends relatively strongly on Mn concentration~\cite{furdy;jap87}.  That the total initial asymmetry $a_1 + a_2$ is only $\sim$0.05 at 2~K compared to 0.22 at 88~K confirms that the transition occurs in the bulk of the sample, rather than some smaller volume fraction.  The higher onset temperature of the spin freezing relative to expectations from the dc magnetometry data can be attributed to the fact that \muSR\ is sensitive to spin fluctuations on a faster timescale than dc magnetometry~\cite{hilli;nrmp22}, so fluctuations just above \Tf\ are slow enough to appear static for \muSR\ but still dynamic for magnetometry. Similar behavior has been observed in \muSR\ data collected on magnetic nanoparticles exhibiting superparamagnetism and blocking behavior~\cite{frand;prm21}, although we note that there are important differences between the dynamics of spin glasses and noninteracting magnetic nanoparticles~\cite{jonss;chapter;acp04}.
	
	A strong temperature dependence is also observed in the relaxation rates, particularly $\lambda_1$. As the temperature decreases toward \Tf\ from above, $\lambda_1$ increases steadily, forming a sharp peak around 24~K before dropping again as the temperature is lowered further. The peaked temperature dependence of the relaxation rate is a classic signature of a critical slowing down of spin fluctuations as a transition is approached, and the location of the peak around 24~K marks \Tf, in good agreement with the dc magnetometry data.
	
	We also performed \muSR\ measurements at 2~K and 88~K in an applied longitudinal field (LF) up to 0.4~T. The LF spectra (data not shown) at 88~K show minimal recovery (i.e. decoupling) of the asymmetry with increasing applied field, ruling out the possibility that weak static internal fields cause the observed relaxation and instead confirming that both the fast and slow relaxation channels originate from dynamic spin fluctuations. In contrast, the LF spectra collected at 2~K do exhibit significant decoupling, confirming the presence of static magnetic moments at low temperature.
	
	The observed \muSR\ behavior confirms a bulk transition around 24~K, consistent with the expected spin-glass transition, but important details of the \muSR\ data differ sharply from behavior seen in canonical spin glasses. In both dilute and concentrated canonical spin glasses, the \muSR\ spectra can typically be modeled by a single-component mathematical function, often a ``stretched exponential'' of the form $a(t)\sim e^{-(\lambda t)^{\beta }}$ or a stretched exponential multiplied by the Kubo-Toyabe lineshape~\cite{uemur;prb85,campb;prl94}. In the present case, such a single-component function is unable to provide a reasonable fit to the data; two components are necessary. Furthermore, the ``stretching'' factor $\beta$ converges to unity for both components in the present case; in canonical spin glasses, the observed value above \Tf\ is 0.5 in dilute spin glasses~\cite{uemur;prb85} and ranges from unity at high temperature to 0.33 at \Tf\ in concentrated spin glasses~\cite{campb;prl94}.
	
	On the other hand, very similar asymmetry spectra consisting of two simple exponential components have been observed in cluster spin glasses~\cite{satoo;hfi97, kawas;jpsj16}. In this picture, then, the fast- and slow-relaxing components of the asymmetry would arise from muons stopping within (or very near to) a spin-correlated cluster (roughly 80\% of the muons based on the value of $a_1$) or muons stopping instead near spins behaving as isolated paramagnetic moments (about 20\% of the muons).
	
	To assess the validity of the cluster spin glass scenario, we estimate the typical correlation time $\tau_c$ from the \muSR\ asymmetry relaxation rates via the well-known relation~\cite{uemur;prb85} $\tau_c \approx \lambda / \Delta^2$, where $\lambda$ is the observed relaxation rate and $\Delta=\gamma_{\mu}B_i$ is the product of the muon gyromagnetic ratio $\gamma_{\mu}$ and the root-mean-square average of the internal field $B_i$ at the muon site. Based on the quasi-instantaneous relaxation of the majority of the \muSR\ asymmetry within the first $\sim$0.01~$\mu$s observed at low temperature, we take $\Delta\approx100$~$\mu$s$^{-1}$ as a reasonable estimate. Plugging in the values of $\lambda_1$ determined from the fits, we estimate $\tau_c$ to be approximately $10^{-10}$~s for $T \gg T_f$. This spin correlation time is much longer than typical dilute spin glasses, for which $\tau_c$ is usually around $10^{-13}$~s~\cite{uemur;prb85}, but is on par with representative cluster spin glasses with $\tau_c$ in the range $10^{-7}$ to $10^{-10}$~s~\cite{roych;aipa23}. This time scale is consistent with a Vogel-Fulcher analysis of previously published ac magnetometry data~\cite{shand;prb98}, which we show in the Supplemental Material.  
	
	One additional point about the \muSR\ data bears mentioning. The temperature dependence of $\lambda_1$ above \Tf\ appears to have a similar power-law shape observed in many concentrated spin glasses. To test this, we took the approach established in the \muSR\ literature~\cite{telli;jmmm95, stewa;prb99, telli;pb00} and fit $\lambda_1$ as a function of temperature $T$ according to the equation
	\begin{equation}
		\label{eq:lambda}
		\lambda(T) = \Lambda_0\left(\frac{T-T_f}{T_f}\right)^{\nu},
	\end{equation}
	where $\Lambda_0$ is an intrinsic relaxation rate and $\nu$ is the critical exponent. The fit is displayed as a log-log plot in the insert of Fig.~\ref{fig:musr}(a), where the highly linear trend confirms the power law behavior. The best-fit values are $\Lambda_0 = 9.2(6)$~$\mu$s$^{-1}$, $\nu = -0.59(6)$, and $T_f = 23.9(7)$~K, where the latter agrees well with the dc magnetometry data. The critical exponent $\sim-0.59$ is a bit smaller in magnitude than typical values between $-0.9$ and $-1.4$ reported in the literature for concentrated spin glasses~\cite{telli;jmmm95, stewa;prb99, telli;pb00}, which may be attributable to the more cluster-like behavior of \znmnte.


	
	\subsection{Neutron Diffraction}
\begin{figure}
		\centering
		\includegraphics[width=80mm]{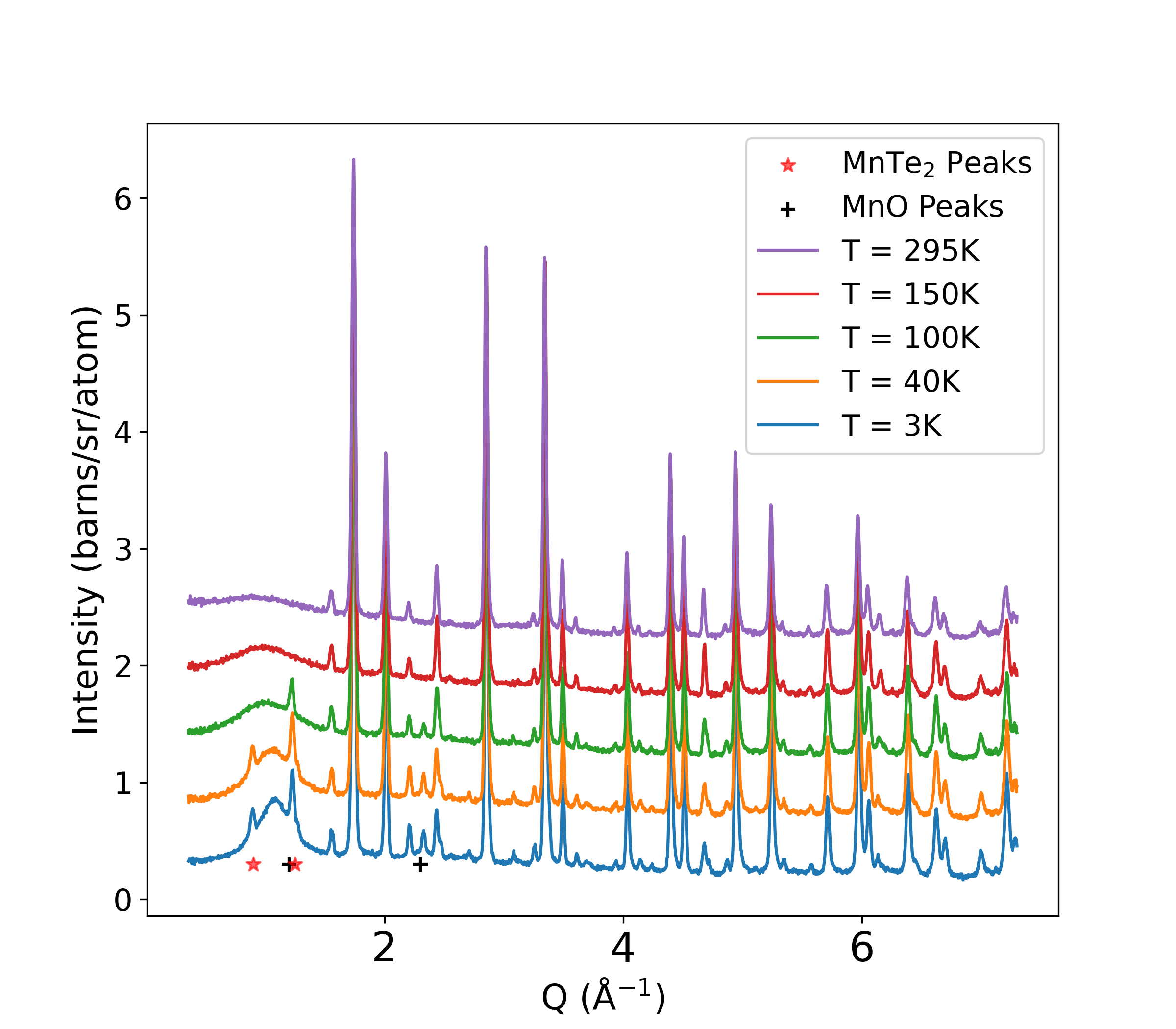} 
		\caption {Representative neutron powder diffraction patterns for \znmnte\ at various temperatures spanning the spin-glass freezing temperature $T_f \approx 23$~K (only visually distinct data sets shown). A prominent hump of diffuse magnetic scattering is visible around 1.1~\AA$^{-1}$, which persists in diminished form to the highest temperature measured. The asterisks and plus signs mark magnetic Bragg peaks from the MnTe$_2$ and MnO impurity phases, respectively. } 
		\label{fig:tempSeries}
	\end{figure}

	Neutron diffraction probes spin-pair correlations directly, providing an opportunity to gain detailed knowledge of the local magnetic structure of \znmnte. In Fig.~\ref{fig:tempSeries}, we display representative neutron powder diffraction patterns collected at several temperatures, normalized to barns per steradian per atom (details given in the Supplemental Material).
	In addition to the sharp nuclear Bragg peaks from the crystal structure of \znmnte, a broad hump of diffuse scattering centered around 1.1~\AA$^{-1}$ is clearly visible at low temperature, indicative of short-range spin correlations in the spin-glass state (the magnetic origin of the diffuse scattering will be argued in the next section). A few sharp Bragg peaks that disappear at higher temperature are also present at low $Q$ (marked by $*$ and $+$), but based on their positions and temperature dependence, these can be identified as AFM peaks from the MnTe$_2$~\cite{burle;prb97} and MnO~\cite{pask;prb01} impurities.
	
	Returning to the diffuse scattering feature, we note that with increasing temperature, the diffuse hump becomes both weaker in amplitude and broader. This indicates a lower overall level of orientational correlations between spins and a shorter correlation length. However, this evolution occurs gradually, with no drastic changes evident at or near $T_f \approx 23$~K. Importantly, the neutron diffraction patterns represent the energy-integrated differential scattering cross section, which probes the \textit{instantaneous} (rather than time-averaged) spin correlations in the material.  This holds as long as the quasistatic approximation~\cite{booth;b;pnscm20} is valid, which is a reasonable approximation in the present case with the incident neutron energy of 34.6~meV and the energy range of magnetic fluctuations extending only up to about 18~meV, and the majority of the spectral weight at still smaller energies~\cite{giebu;prb89, henni;prb02} . The persistence of magnetic diffuse scattering well above \Tf\ demonstrates that dynamically correlated spin fluctuations remain in the paramagnetic state but are time-averaged away to zero by slower probes such as magnetometry. This is likewise consistent with the \muSR\ data. 
	
	\subsubsection{Data Treatment to Isolate Diffuse Magnetic Scattering}
	
	
	To gain deeper insights into the local magnetic structure of \znmnte\ through mPDF analysis and RMC modeling, it is necessary to separate the magnetic contributions to the neutron scattering data from the nuclear contributions. For our particular data set, the most common methods for accomplishing this separation were unsuccessful or could not be applied, requiring a new approach that we describe subsequently. In the ideal case, polarized neutrons could be used to separate the magnetic and nuclear scattering unambiguously, as has previously been done for mPDF analysis~\cite{frand;jap22}, but the experiment in our study was performed with unpolarized neutrons. Another common approach is to collect a diffraction pattern in the purely paramagnetic state, such that the spin orientations are completely random and no structured diffuse scattering exists, and then subtract this high-temperature pattern from the lower-temperature patterns to remove the (ideally) temperature-independent nuclear Bragg peaks and isolate the temperature-dependent diffuse magnetic scattering signal. In our case, even the highest-temperature diffraction pattern at room temperature contains appreciable diffuse magnetic scattering, so a straight subtraction would not work. Significant shifts in the peak positions due to thermal expansion would also present challenges for this method, as would the presence of the impurity magnetic Bragg peaks at low temperature.
	
	Another possibility is to fit a structural model to the nuclear component and subtract the resulting calculated nuclear scattering from the data, leaving the magnetic component. In principle, this can be done in either reciprocal space or real space. In practice, this approach can be challenging in reciprocal space, because even relatively small misfits to the data, such as imperfect simulation of the Bragg peak shape, can significantly contaminate the isolated magnetic signal and resulting mPDF pattern~\cite{frand;aca15}. The real-space approach has seen comparatively more success. Computing the Fourier transform of the nuclear and magnetic scattering data together produces the total PDF, i.e. the sum of the atomic and magnetic PDF patterns. The atomic PDF can then be fit to the data and subtracted, leaving only the mPDF. This works well when the accessible $Q$ range of the measurement is large enough to produce an atomic PDF signal that can be modeled in a meaningful way. In our case using 1.54~\AA\ neutrons on HB2A, the diffraction patterns extend only to 8~\AA$^{-1}$. This is adequate for mPDF data, since the magnetic form factor typically suppresses any meaningful signal beyond this value, but it is insufficient for generating the atomic PDF. If the total PDF were generated using diffraction patterns with this $Q$ range, the truncation artifacts would be too severe to allow proper removal of the atomic PDF. The shorter wavelength options at HB2A provide a larger $Q$-range such that this real-space separation of the nuclear and magnetic signals can be accomplished~\cite{baral;prb24}, and we found some success with this approach using the single data set collected at 3~K with 1.12~\AA\ neutrons (see Supplemental Material). However, the rest of the data collected with the longer wavelength option (which tends to yield a cleaner magnetic signal) require a different treatment.

	Considering the challenges faced by established methods for our data, we implemented a novel algorithm to detect and remove Bragg peaks automatically from powder data sets, leaving only the diffuse scattering. In this algorithm, the standard deviation of the scattered intensity is computed within a sliding window that passes continuously through the entire data range, yielding an array of standard deviations as a function of $Q$. Peaks are then identified as regions of standard deviation above a user-defined threshold. All regions exceeding the threshold are excised from the original scattering pattern, and linear interpolation is used to fill the resulting gaps. The resulting diffuse scattering pattern is somewhat choppy, so convolution with a narrow Gaussian is used to remove abrupt kinks near peak edges. For the case of \znmnte{}, we chose the width of the window for computing the standard deviation to be 0.056 \AA$^{-1}$ after some trial and error. Further, we found that it was suboptimal to use a constant threshold for identifying peaks, because the standard deviation within the diffuse scattering region was larger than that of some of the small Bragg peaks in the mid-$Q$ range. If a $Q$-dependent threshold is used, it should be smoothly varying and contain as few parameters as possible to minimize any bias that could obscure real features. To account for both concerns,  we defined the threshold as a shallow quadratic function, which effectively selected the diffuse scattering over the whole $Q$ range.  
	
\begin{figure}
		\centering
		\includegraphics[width=8.7cm]{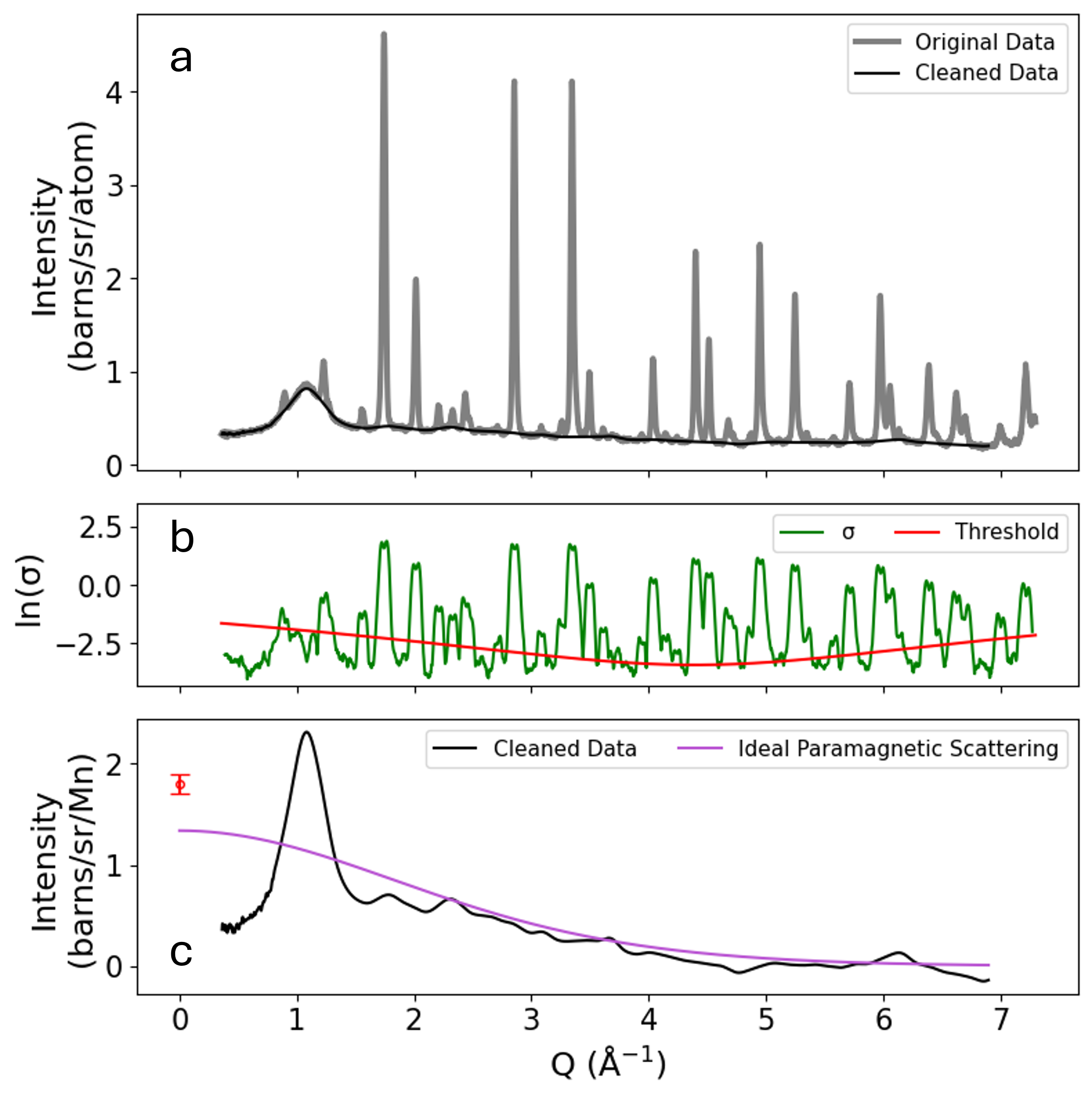}%
		\caption{Extraction of the diffuse scattering signal from the diffraction pattern at 3~K. (a) Cleaned data overlaid on raw (background subtracted) data. (b) Logarithm of the standard deviation, overlaid by the threshold used to determine the position of peaks, as explained in the main text. (c) Cleaned data (less the high-$Q$ constant asymptote) overlaid with the ideal paramagnetic scattering, showing that the diffuse scattering follows the expected trend nicely. The purple curve was calculated assuming $S=2.18$ as determined from the RMC fits described in the main text. The red symbol marks the expected paramagnetic scattering at $Q=0$ based on the Curie-Weiss fits to the susceptibility.}%
		\label{fig:datacleaning}%
	\end{figure}

	
	The result of applying this algorithm to the \znmnte\ data collected at 3~K is shown in Fig.~\ref{fig:datacleaning}.
	The algorithm successfully removed all Bragg peaks, leaving a relatively clean diffuse scattering signal suitable for further analysis. The algorithm was similarly effective for diffraction patterns collected at all other temperatures. We note that excessive peak overlap in the scattering pattern could pose challenges for this algorithm, but considering the high symmetry of \znmnte, it was not an obstacle here. Finally, we mention that our python implementation of this algorithm also includes an option to add any magnetic Bragg peaks selected by the user back into the scattering pattern after their initial removal. We did not use this feature in the current study, since no magnetic Bragg peaks exist for \znmnte, but it could potentially be useful for situations of coexisting magnetic Bragg peaks and magnetic diffuse scattering.
	
	With the diffuse scattering isolated, it is crucial to determine whether it is purely magnetic or whether contributions from chemical short-range order could also be present. We argue that the diffuse scattering in \znmnte\ is magnetic in origin, with no or negligible nuclear contributions, based on the following observations. First, the diffuse scattering in Fig.~\ref{fig:tempSeries} shows a strong temperature dependence, consistent with magnetic scattering. Second, the magnitude and $Q$-dependence of the diffuse scattering closely match the scattering for an ideal paramagnet (i.e., randomly oriented spins),
	\begin{equation}
		\frac{1}{N_s}\left(\frac{\mathrm{d}\sigma}{\mathrm{d}\Omega}\right)_{\mathrm{paramagnet}}=\frac{2}{3}\left(\gamma r_0\right)^2S(S+1)\left[f_m(Q)\right]^2,
	\end{equation}
	where the symbols are as defined in Eq.~\ref{eq:FT}. The observed scattering is especially similar to the ideal paramagnetic scattering at 295~K where the magnetic correlations are weak (see Supplemental Material), but the similarity is also present at 3~K, as seen in Fig.~\ref{fig:datacleaning}(c). As an independent check of the overall scale of the diffuse scattering, Fig.~\ref{fig:datacleaning}(c) also shows the $Q=0$ value of the paramagnetic intensity corresponding to the effective moment size estimated from the Curie-Weiss fits to the susceptibility, which is slightly higher than expected, but still generally consistent. Third, the X-ray PDF analysis reveals no evidence for features of the local structure that hint at chemical short-range ordering, as discussed previously. The magnetic origin of the diffuse scattering underlies numerous other neutron scattering studies of (Zn,Mn)Te and related compounds~\cite{holde;prb82, furdy;jap87, furdy;jap88}; the arguments presented here further support that historical position. Additionally, we consider any diffuse magnetic scattering from the impurity phases MnTe$_2$ and MnO above their respective N\'eel temperatures to be negligible, since the molar fraction of these combined impurity phases is below 5\%. We note that MnO is known to have prominent magnetic diffuse scattering above \TN~\cite{shull;pr51}, but MnO data collected on a similar neutron diffractometer by a subset of the authors for previous studies~\cite{frand;aca15,frand;prl16} suggest that the diffuse intensity should be about 50 times weaker than the magnetic Bragg peak intensity for MnO, safely negligible in the current study.

	\subsection{Magnetic PDF analysis}
	
\begin{figure}
		\centering    \includegraphics[width=85mm]{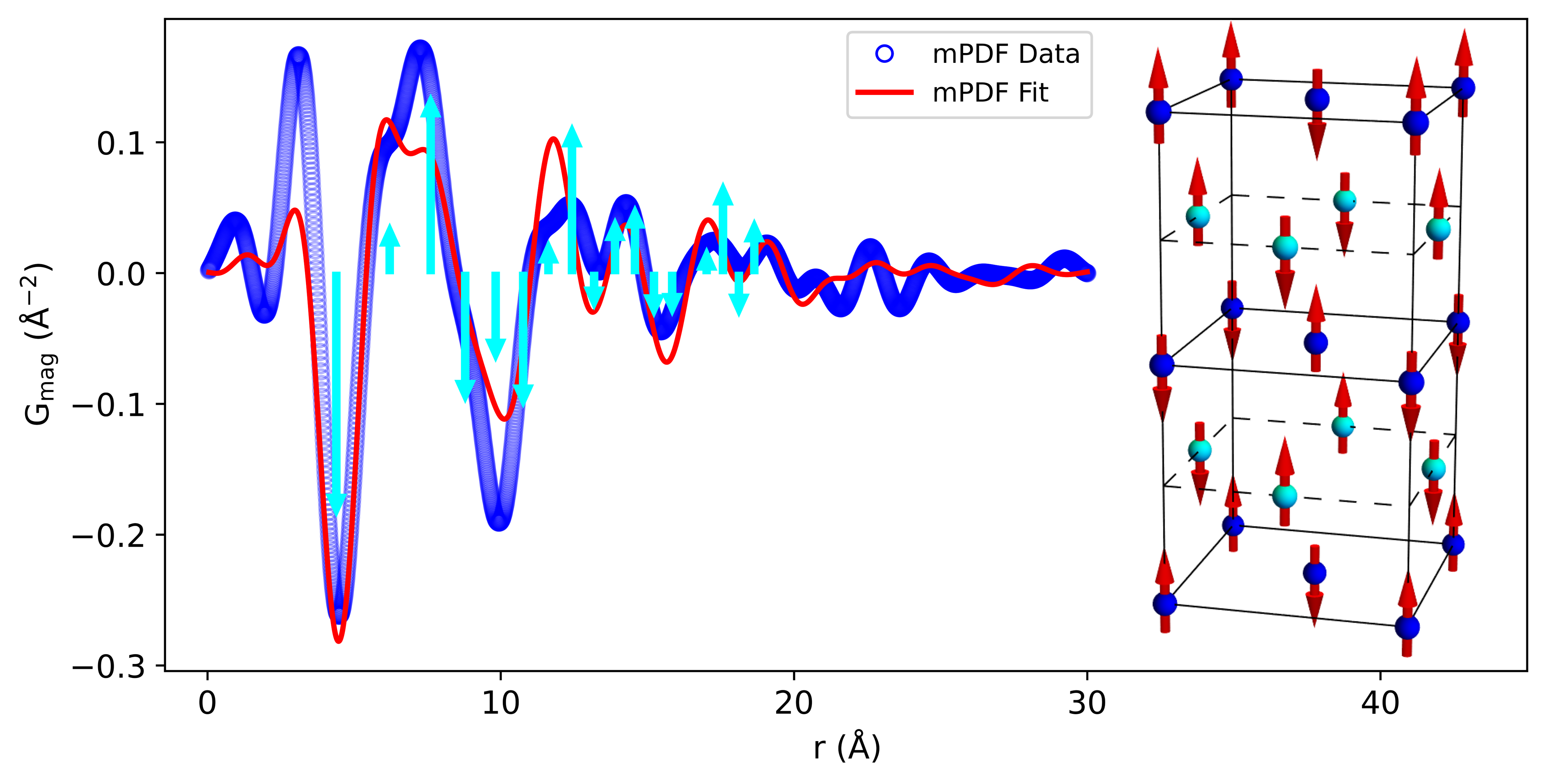}  
		\caption {Fit of antiferromagnetic type-III structure to experimental deconvolved magnetic pair distribution function (mPDF) for \znmnte\ at 3~K. The average correlation function (calculated over 100 Spinvert fits) is shown as cyan arrows overlaid on the data. Inset: Antiferromagnetic type-III order, with the propagation vector along the vertical axis.}
		\label{fig:mPDFfit}
	\end{figure}

	To continue our analysis of the short-range magnetic correlations in \znmnte, we now turn to the mPDF, which we generated for each temperature by Fourier transforming the isolated diffuse magnetic scattering. The blue curve in Fig.~\ref{fig:mPDFfit} shows the mPDF at 3~K, labeled as $G_{\mathrm{mag}}$ and deconvolved from the effect of the magnetic form factor for improved real-space resolution. The local magnetic structure is immediately evident in the mPDF pattern. We observe a large, negative peak at the nearest neighbor distance of $\approx$4.4~\AA, indicative of strong antiferromagnetic nearest-neighbor correlations. The peaks below this distance are truncation artifacts from the Fourier transform. Two positive peaks are visible at distances of approximately 6.2~\AA\ and 7.6~\AA, corresponding to net parallel alignment of second- and third-nearest neighbors. Negative as well as positive peaks continue to appear at further interatomic distances, reflective of the net antiferromagnetic local spin arrangement. The magnitudes of these peaks diminish steadily with increasing $r$, demonstrating the short-range-ordered nature of the spin-glass state and indicating that non-random spin correlations persist on the length scale of about 2~nm at 3~K.
	
	
	
	We performed a more quantitative analysis of the mPDF data by fitting various magnetic structure models to the data. Most antiferromagnets with the fcc structure exhibit one of a few collinear magnetic configurations, often referred to as types I, II, and III~\cite{chatt;06b}. We modeled the data with each of these three structures, using least-squares optimization to refine a correlation length, the spin direction, and the size of the locally ordered magnetic moment per Mn$^{2+}$ ion at the nearest-neighbor distance, which we refer to as the local magnetic order parameter (LMOP)~\cite{zhang;prb19}. For a collinear model, the LMOP can be quantified as $\sqrt{|\langle \bm{\mu}_i\cdot \bm{\mu}_j \rangle|}$, where $\bm{\mu}_i$ and $\bm{\mu}_j$ are the magnetic moments associated with nearest-neighbor spins $\mathbf{S}_i$ and $\mathbf{S}_j$, and the average in angled brackets is taken over all nearest-neighbor pairs of magnetic moments. Thus, the LMOP is zero in the case of a perfectly random configuration of spins and has a maximum of $2S\mu_B$ (5~\muB\ in the present case, assuming $S=5/2$) for maximally aligned or anti-aligned nearest-neighbor spins. It is useful because it quantifies the strength of the instantaneous local magnetic correlations (at the nearest-neighbor distance) even when no long-range order is present. 
	
	Only the type-III order (illustrated in the inset of Fig.~\ref{fig:mPDFfit}) provided a reasonable fit to the data, shown as the red curve in Fig.~\ref{fig:mPDFfit}.
	This result confirms a previous prediction of short-range type-III order in this class of materials~\cite{holde;prb82}, while also providing greater detail than is available from earlier studies. Specifically, we found that the LMOP is 3.1(1)~\muB, with the correlation strength decaying exponentially in $r$ as $\exp(-r/\xi)$ with the correlation length $\xi = 5.7(1)$~\AA. As an additional check on the magnitude of the LMOP, we used the scale factor for the atomic PDF generated from the data collected with 1.12~\AA\ neutrons to calibrate the strength of the magnetic PDF~\cite{fletc;prb21}, as explained in the Supplemental Material. We arrived at 3.2(3)~\muB, consistent within uncertainty.  The locally correlated moment of 3.1~\muB\ is smaller than the 5~\muB\ expected for fully ordered $S=5/2$ spins, which could be due to residual frozen-in disorder, some fraction of Mn$^{2+}$ spins freezing into completely random orientations (e.g. spins that do not happen to have any other spins within their first or second coordination shell, such that they experience no strong exchange interactions) and/or some reduction of the bare local moment from the maximum allowed value due to orbital hybridization. The latter is consistent with the optimized value of $S = 2.18$ that was found from the RMC analysis described in the next section.~The best-fit spin direction is approximately along the propagation vector $(0, 1/2, 0)$, similar to the classic type-III fcc antiferromagnet MnS$_2$~\cite{corli;jap58}. Regarding the possibility of coexisting type-I and type-III correlations proposed in Ref.~\onlinecite{ono;jpcs99}, we find no evidence for any type-I correlations in the neutron diffraction or mPDF data. Adding a fitting component with short-range-ordered type-I correlations to the mPDF fits provides no appreciable improvement to the fit quality; furthermore, the diffuse scattering pattern has no hint of excess intensity centered around the $Q=(100)$ position that would correspond to type-I short-range order. 
	
\begin{figure}
		\centering    \includegraphics[width=80mm]{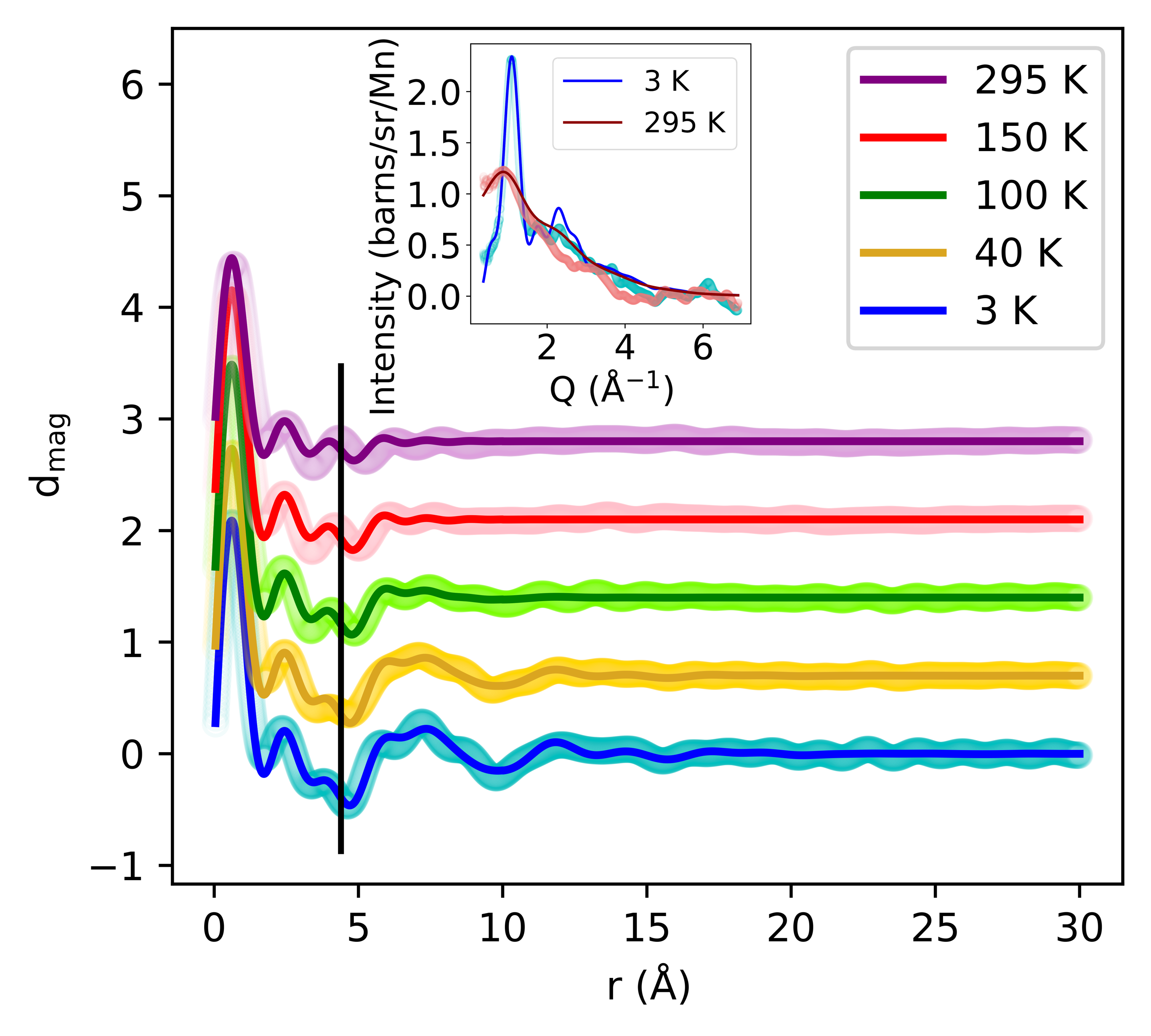}  
		\caption {Non-deconvolved mPDF $d_{\mathrm{mag}}$ at representative temperatures between 3~K and 295~K (same temperatures shown in Fig.\ref{fig:tempSeries}), offset vertically for clarity. Fits using the antiferromagnetic type-III model are overlaid as the darker solid curves. The solid vertical line indicates the nearest neighbor distance, below which Fourier artifacts often accumulate. The inset shows the calculated magnetic scattering corresponding to the mPDF fits at 3~K and 295~K (solid curves) compared to the observed diffuse magnetic scattering (open circles).}
		\label{fig:tempDmag}
	\end{figure}

	We performed similar fits to the mPDF data collected at all other temperatures, as well. Due to the weaker diffuse scattering signal at higher temperatures, attempting to deconvolve the diffuse scattering from the magnetic form factor prior to Fourier transformation led to untenable levels of noise and Fourier artifacts in the data. Accordingly, we opted instead to use the non-deconvolved mPDF data, which we label $d_{\mathrm{mag}}$. This has the advantage of reducing the noise in the mPDF data, but at the cost of reduced real-space resolution. Fig.~\ref{fig:tempDmag} shows representative $d_{\mathrm{mag}}$ data and fits.
	We note that the large positive peak below 1~\AA\ is unrelated to any spin-spin correlations and is simply a byproduct of the procedure used to generate the non-deconvolved mPDF, as explained elsewhere~\cite{frand;aca15}. The fit at 3~K is quite good and yields similar results as the fit to the deconvolved mPDF $G_{\mathrm{mag}}$ at that temperature, except the correlation length refines to 4.1(1)~\AA, somewhat shorter than the value obtained from the $G_{\mathrm{mag}}$ fit. This may be due to the broadening of weak mPDF features at longer $r$ such that they are indistinguishable from the background, shortening the $r$-range over which meaningful signal exists relative to the deconvolved $G_{\mathrm{mag}}$. We emphasize that the correlation length is simply the characteristic decay length of the spin correlation function, and not the maximum distance for which meaningful spin correlations are present. A correlation length shorter than the nearest-neighbor distance has no special significance other than indicating that the magnitude of the spin-spin correlations decreases rapidly with distance, so meaningful correlations beyond the nearest neighbors would only be expected if they fall within a few correlation lengths of the nearest-neighbor distance.
	
	Important information can be gleaned from the mPDF data displayed in Fig.~\ref{fig:tempDmag}. First, the mPDF remains qualitatively unchanged between 3~K and 40~K, indicating that the instantaneous local magnetic structure does not undergo a significant change across \Tf. In that sense, this neutron experiment is essentially blind to the spin-glass transition, since this energy-integrated neutron scattering measurement probes the instantaneous spin correlations, which evidently do not change at the freezing temperature. This aspect of neutron diffraction data on spin glasses has been discussed elsewhere~\cite{fisch;b;sg93}. Neutron spin echo measurements would be a more effective neutron-based probe of the spin-glass dynamics. 
	
\begin{figure}
		\centering    \includegraphics[width=80mm]{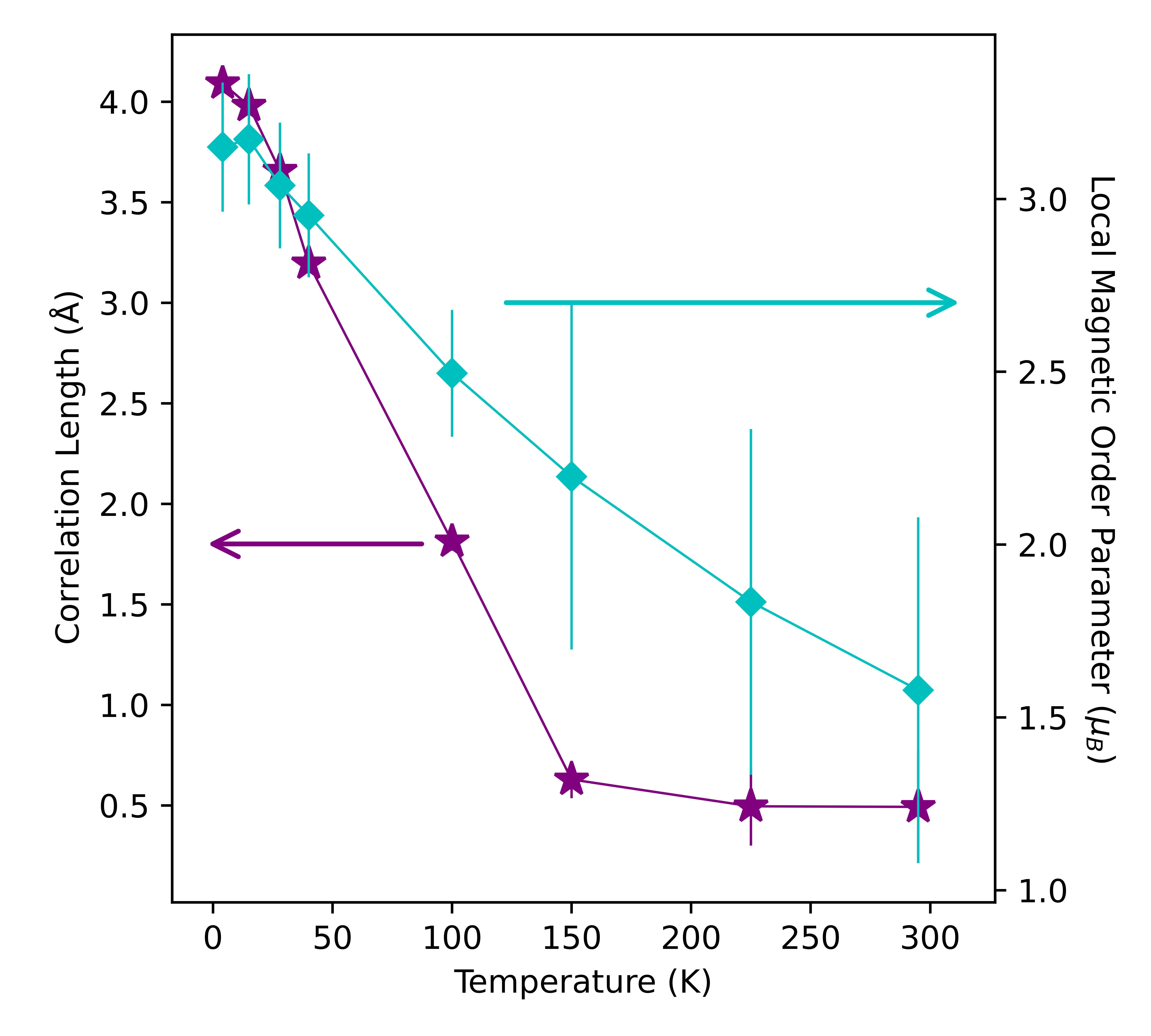}  
		\caption {Temperature dependence of the correlation length (left vertical axis) and locally ordered moment (right vertical axis) obtained from fits to $d_{\mathrm{mag}}$.}
		\label{fig:tempTrends}
	\end{figure}

	The second important feature of the temperature-dependent mPDF data is that the amplitude of the mPDF decreases with increasing temperature, pointing to increased randomness between neighboring spins and a reduction in the LMOP. Third, the $r$-range over which meaningful oscillations in the mPDF signal exist becomes successively shorter with increasing temperature, indicating that the magnetic correlation length decreases. At 295~K, only a slight negative dip around the nearest-neighbor distance is visible (although somewhat obscured by the Fourier ripples), indicating that nearest-neighbor antiferromagnetic correlations persist at room temperature but do not extend to further distances. This analysis is consistent with the modulation of the diffuse magnetic scattering pattern on top of the squared magnetic form factor observed at 295~K, displayed in the inset of Fig.~\ref{fig:tempDmag} along with the diffuse scattering at 3~K for comparison. The best-fit correlation length and local magnetic order parameter values for each temperature are displayed in Fig.~\ref{fig:tempTrends}, confirming the downward trend with temperature inferred from qualitative inspection of the data.
	While it may seem surprising that nearest-neighbor correlations could survive to such a high temperature relative to \Tf, this can be understood by the relatively strong exchange interaction of $\sim$8 -- 9~K determined via inelastic neutron scattering~\cite{corli;prb86,giebu;prb89}, equating to an interaction energy $J \mathbf{S}\cdot\mathbf{S}$ around 50~K. Since non-random spin correlations commonly persist to temperatures several times the interaction energy~\cite{paddi;jpcm13,frand;cm24}, it is reasonable that nearest-neighbor AFM correlations should remain up to at least room temperature in \znmnte. Related to this is the Curie-Weiss temperature, which is approximately $-400$~K for \znmnte~\cite{mcali;prb84}, further establishing the large energy scale of the magnetic interactions and the expectation of non-random magnetic correlations to similarly high temperatures. 
	
	\subsection{Reverse Monte Carlo Modeling}
	
\begin{figure}
		\centering
		\includegraphics[width=8cm]{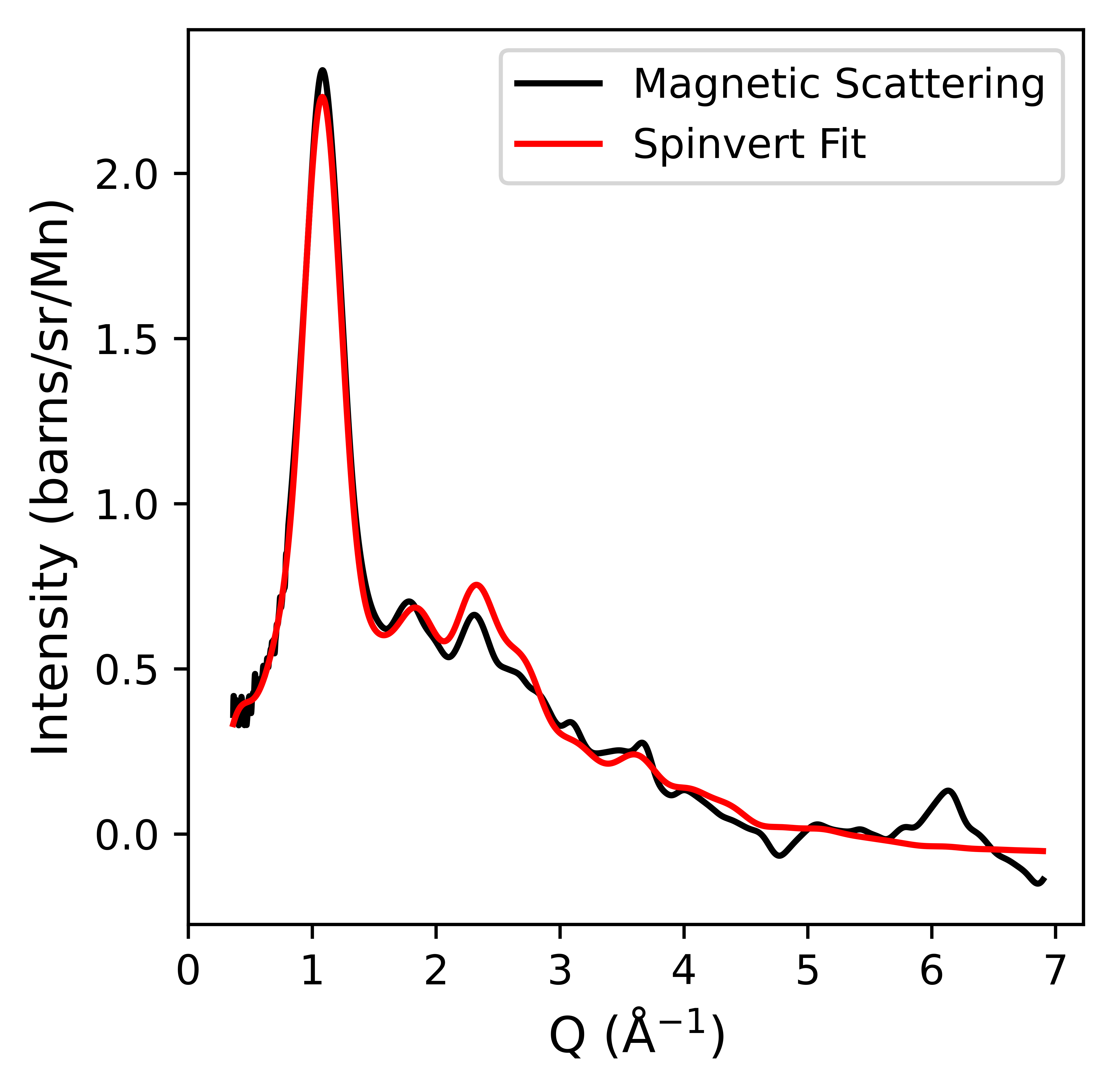}%
		\caption{RMC modeling of magnetic diffuse scattering data for \znmnte\ at 3~K, as carried out in Spinvert.}
		\label{fig:spinvertRegular}%
	\end{figure}

	
	We next present RMC fits to the diffuse magnetic scattering data at 3~K, providing an independent consistency check for our findings from the mPDF analysis.  The two approaches are highly complementary: Viewing the data in real space can offer a more intuitive understanding of the local magnetic correlations and simpler evaluation of successes and shortcomings of candidate models, while reciprocal space may offer a clearer way to distinguish real features from artifacts; fitting to the mPDF using a well-controlled, small-box model with just a few free parameters enables a more straightforward interpretation of results with reduced likelihood of fitting to spurious features of the data, but large-box reverse Monte Carlo modeling can reveal important features without relying on an assumed model. Taking an all-of-the-above approach~\cite{nutta;prb23} provides greater confidence in the results that are found to be self-consistent. To carry out the RMC refinements,we defined a $6\times6\times6$ supercell and performed 100 RMC fits to build up a statistical distribution of spin configurations consistent with the data. The average of these fits is shown in Fig. \ref{fig:spinvertRegular}, demonstrating good agreement between the model and the data. Allowing the Spinvert program to optimize the scale factor resulted in an effective spin size of $S = 2.18$, a bit smaller than the maximum value $S=5/2$ expected for bare Mn$^{2+}$ moments, but in close enough agreement to validate the normalization of the scattering data to absolute units.

	We calculated the spin correlation function as defined in Eq.~\ref{eq:spincorrelation} for each of these fits, and then averaged all 100 to produce a final spin correlation function. We display this function, which is defined only for distances corresponding to Mn-Mn pairs, as a series of cyan arrows in Fig.~\ref{fig:mPDFfit}. The arrows represent the magnitude and net orientation (positive or negative) of the average correlation function at each Mn-Mn separation distance. The spin correlation function visualized in this way follows the shape of the mPDF data and fit quite well, confirming that the local magnetic correlations exhibit strong type-III character.  The spin correlation length extracted by fitting an exponential function to the magnitude of the spin correlation function~\cite{paddi;jpcm17} is 7(2)~\AA\ at 3~K, consistent with the 5.7(1)~\AA\ correlation length determined from fits to the deconvoluted mPDF. The magnitude of the spin correlation function at the nearest neighbor distance (i.e. the LMOP) equates to 2.3~\muB, generally consistent with, but somewhat less than, the 3.1(1)~\muB\ determined by mPDF analysis. This is not surprising, since RMC is known to systematically underestimate the locally ordered moment~\cite{paddi;jpcm13}.
	
	\section{Discussion and Conclusion}
	
	
	The results presented in this work constitute compelling evidence for the cluster spin glass scenario proposed for (Zn,Mn)Te~\cite{furdy;jap87} and provide important new details regarding the nature of the spin correlations and dynamics in this system. The cluster-spin-glass ground state in \znmnte\ is characterized by short-range-ordered type-III AFM clusters that extend over approximately 2~nm with a characteristic spin correlation decay length of 4-5~\AA. The predominance of type-III correlations in \znmnte\ is consistent with the observation of long-range type-III order in thin films of zincblende MnTe~\cite{kloso;jap91, henni;prb02}; indeed, the local magnetic correlations in bulk \znmnte\ can be well described as a short-range version of the magnetic order in thin-film zincblende MnTe (but they bear no resemblance to bulk MnTe, which crystallizes in a hexagonal structure and has a completely different AFM structure~\cite{kunit;jdp64}). The additional disorder caused by Zn/Mn mixing and the geometrical frustration of AFM interactions on the fcc lattice cause a relatively short correlation length and hinder the formation of long-range order in \znmnte. 
	
	As seen in Fig.~\ref{fig:tempTrends}, the LMOP and correlation length are saturated or nearly saturated below \Tf. This demonstrates that the AFM clusters are essentially locked into a random configuration below \Tf\ but retain well-defined intra-cluster type-III correlations. With nonexistent or exceedingly slow spin dynamics below \Tf, type-III alignment is unable to strengthen within any given cluster (thus limiting the LMOP) and neighboring clusters and/or uncorrelated spins near a cluster are unable to come into alignment, which limits the correlation length. Above \Tf, the AFM clusters begin to fluctuate---very slowly just above \Tf, but with the average time between fluctuations decreasing to about 10$^{-10}$~s at 88~K based on our \muSR\ analysis, a typical correlation time for cluster-spin-glass systems~\cite{roych;aipa23}. Importantly, the mPDF pattern undergoes no qualitative change just above \Tf. This demonstrates that the intra-cluster type-III spin correlations are preserved despite the fluctuations, implying that the clusters fluctuate as individual units. The type-III correlations extend over at least the first two or three coordination shells up to about 100~K, but at higher temperatures, the clusters dissociate into liquid-like fluctuations of AFM-correlated nearest-neighbor spin pairs. The relatively broad temperature interval above \Tf\ over which the initial \muSR\ asymmetry is lost [Fig.~\ref{fig:musr}(a)] indicates that this dynamical evolution occurs somewhat inhomogeneously throughout the sample volume; whether this is an intrinsic feature of concentrated spin glasses or a consequence of random spatial variations of Mn concentration in the sample is an open question.
	
	
	The success of this study in providing essential new details into the spin-glass correlations and dynamics of \znmnte\ via modern \muSR\ and neutron experimentation and data analysis encourages future studies of systems with short-range magnetism, even classic systems such as (Zn,Mn)Te, using a similar approach as employed here. In addition, this work introduces valuable technical developments for mPDF and diffuse scattering studies conducted on neutron instruments that are not specifically optimized for such experiments. In particular, the data-cleaning algorithm to remove Bragg peaks and isolate diffuse scattering signals allows effective mPDF analysis even when the preferred methods, such as using polarized neutrons, subtracting a high-temperature purely paramagnetic reference pattern, or~generating the combined atomic and magnetic PDF by Fourier transforming both the nuclear and magnetic scattering together, cannot be used. These developments may significantly expand the range of experiments for which mPDF can be successfully applied. 
	\\
	
	\textbf{Acknowledgements}
	We thank the staff at TRIUMF and the High Flux Isotope Reactor for their valuable help and support during the experiments. The neutron and muon work was supported by the U.S. Department of Energy, Office of Science, Basic Energy Sciences (DOE-BES) through Award No. DE-SC0021134. A portion of this research used resources at the High Flux Isotope Reactor, a DOE Office of Science User Facility operated by the Oak Ridge National Laboratory. The beam time was allocated to HB2A on proposal number IPTS-29767.

\end{document}


\title{Supplemental Material: Cluster spin glass correlations and dynamics in Zn$_{0.5}$Mn$_{0.5}$Te
}
	
	\author{Sabrina R. Hatt}
	\affiliation{ %
		Department of Physics and Astronomy, Brigham Young University, Provo, Utah 84602, USA.
	} %

	\author{Camille Shaw}
	\affiliation{ %
		Department of Physics and Astronomy, Brigham Young University, Provo, Utah 84602, USA.
	} %

	\author{Emma Zappala}
	\affiliation{ %
		Department of Physics and Astronomy, Brigham Young University, Provo, Utah 84602, USA.
	} %

	\author{Raju Baral}
	\affiliation{ %
		Department of Physics and Astronomy, Brigham Young University, Provo, Utah 84602, USA.
	} %
    \affiliation{ %
        Neutron Scattering Division, Oak Ridge National Laboratory, Oak Ridge, Tennessee 37831, USA.
    } %

    \author{Stuart Calder}
    \affiliation{ %
        Neutron Scattering Division, Oak Ridge National Laboratory, Oak Ridge, Tennessee 37831, USA.
    } %

\author{Gerald D. Morris}
\affiliation{Centre for Molecular and Materials Science, TRIUMF, Vancouver, British Columbia, Canada V6T 2A3}

    \author{Brenden R. Ortiz}
    \affiliation{
        Materials Science and Technology Division, Oak Ridge National Laboratory, Oak Ridge, Tennessee 37831, USA.
    }

	\author{Karine Chesnel}
	\affiliation{ %
		Department of Physics and Astronomy, Brigham Young University, Provo, Utah 84602, USA.
	} %

	\author{Benjamin A. Frandsen}
	\affiliation{ %
		Department of Physics and Astronomy, Brigham Young University, Provo, Utah 84602, USA.
	} %

\maketitle
\section{Curie-Weiss fits to magnetic susceptibility data}
\begin{figure}
	\includegraphics[width=90mm]{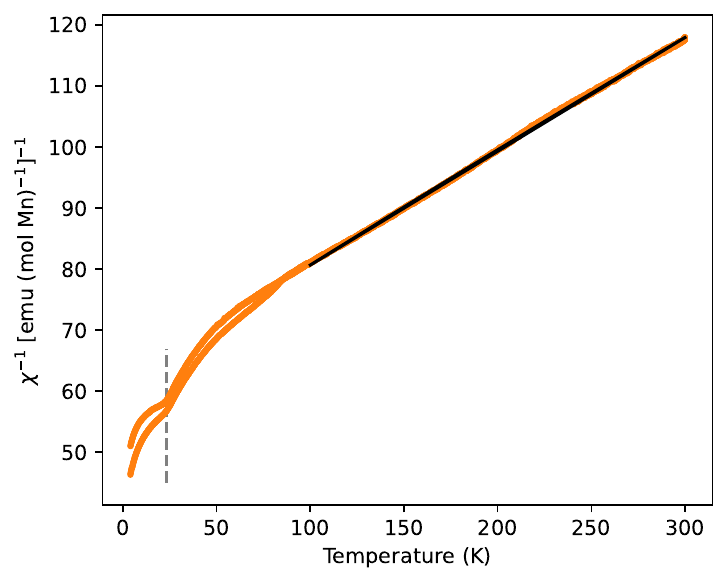}
	\caption{\label{fig:CW} Curie-Weiss fit to the inverse magnetic susceptibility of \znmnte. The orange curve is the data, the black curve the fit, and the vertical dashed gray line marks the freezing temperature.}
\end{figure}

We performed a Curie-Weiss fit to the molar susceptibility data to extract the effective moment size and the Curie-Weiss temperature. Following Mugiraneza and Hallas~\cite{mugir;cph22}, we modeled the inverse susceptibility as 
\begin{equation}
\label{eq:sus}
\chi^{-1} = \frac{T - \theta_{CW}}{\chi_0 \cdot (T - \theta_{CW}) + C},
\end{equation}
where $\theta_{CW}$ is the Curie-Weiss temperature, $C$ is the Curie-Weiss constant related to the effective moment through $\mu_{eff}=\sqrt{8C}\mu_{\mathrm{B}}$ in cgs units, and $\chi_0$ is a small, temperature-independent contribution to the susceptibility that we found necessary to include in the fit to achieve adequate agreement with the data, likely due to the minor impurity phases. The fit over the temperature range 100 - 300~K is shown in Fig.~\ref{fig:CW}. The optimized fitting parameters are $\theta_{CW}=-300(2)$~K, $C=4.70(3)$~emu$\cdot$K/(mol~Mn), and $\chi_0=6.4(3)\times10^{-4}$~emu/(mol~Mn). The Curie-Weiss constant corresponds to an effective moment of $\mu_{eff}=6.13(2)$~\muB, close to the expected value of 5.92~\muB\ for $S=5/2$ spins. Because slightly different values of the effective moment are obtained for different fitting ranges, we estimate a more realistic uncertainty of at least 0.05~\muB\ for $\mu_{eff}$. The intensity of the paramagnetic scattering at $Q=0$ corresponding to this effective moment is given in Fig.~\ref{fig:allpatterns}, showing reasonable agreement with the overall scale observed in the scattering data.

\section{Analysis of previously published ac magnetometry data}
The glassy character of \znmnte{} was examined through previously published AC magnetometry data on a sample with a nearly identical composition~\cite{shand;prb98}. The temperature of the peak in ac susceptibility for each frequency $\nu$, denoted \Tf, was estimated from Fig.~1 in Ref.~\onlinecite{shand;prb98} using WebPlotDigitizer \cite{marin;arxiv17} and fit to the Arrhenius law and the Vogel-Fulcher law, following the method used in \cite{kolay;prb25}. The Arrhenius and Vogel-Fulcher laws are given, respectively, by
\begin{subequations}
\label{eq:arrvf}
\begin{align}
   \tau &= \tau_0 \, \text{exp}\biggl(\frac{E_0}{k_B T_f}\biggr), \label{eq:arrhenius} \\
   \tau &= \tau_0 \, \text{exp}\biggl(\frac{E_0}{k_B (T_f-T_0)}\biggr), \label{eq:vogelfulcher}
\end{align}
\end{subequations}
where $\tau=\frac{1}{2\pi \nu}$ is the relaxation time of the measurement, $\tau_0$ is the intrinsic spin-flip time, $E_0$ is the activation energy of the spin flip, and for the Vogel-Fulcher law, $T_0$ is an empirical parameter related to the interactions between spins.
The Arrhenius law assumes negligible interactions between spins, precluding the possibility of magnetic cluster formation. The Vogel-Fulcher law is a similar model, but it more accurately describes cluster glass interactions through the inclusion of $T_0$.

\begin{figure}
	\includegraphics[width=90mm]{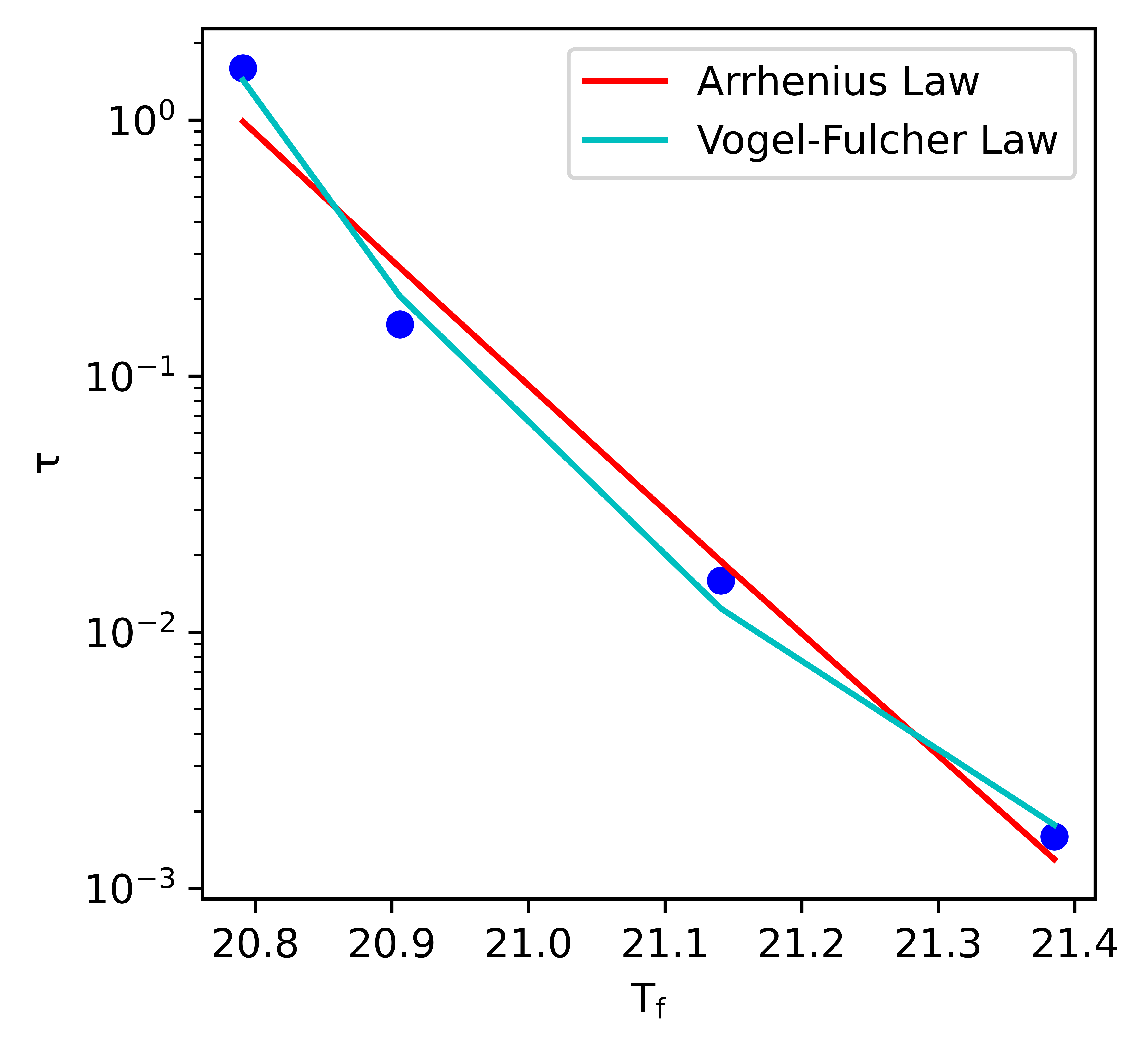}
	\caption{\label{fig:clusterGlass} Fit of Arrhenius and Vogel-Fulcher models to AC magnetometry data. Fits were both performed on a logarithmic scale and appear reasonable, though only the Vogel-Fulcher law gives physically feasible values for the fit parameters. }
\end{figure}

Fig.~\ref{fig:clusterGlass} displays $\tau$ on a log scale versus \Tf, together with the fits using the Arrhenius and Vogel-Fulcher laws. The Arrhenius fit gives unphysical parameter values, casting the assumption of non-interacting spins into doubt: $\tau_0$ is found to be between 10$^{-115}$ and 10$^{-93}$~s, far too short to be physically meaningful, and $E_a/k_B$ comes to 4990 $\pm$ 460~K, far too high to be physical. Using the Vogel-Fulcher model, we obtain $\tau_0$ between 10$^{-10}$ and 10$^{-4}$ seconds, which is consistent with expected time scales for a cluster glass. The activation energy $E_0/k_B$ is also a much more reasonable 160~K, although the statistical uncertainty obtained from the fit is on the same order of magnitude as the parameter value due to the limited number of data points available for the fit, so the precision of this analysis is greatly limited. T$_0$ also refines to a reasonable value of 19.9 $\pm$ 0.6~K. The nonzero value indicates that significant interactions between spins are present, pointing to cluster formation. Overall, analysis of the ac magnetometry data is consistent with expectations for a cluster glass and strongly rules out the possibility of a canonical spin glass with negligible interactions between spins, as seen by the failure of the Arrhenius model.

\section{Checking the absolute normalization of the neutron diffraction intensity}

\begin{figure}
	\includegraphics[width=90mm]{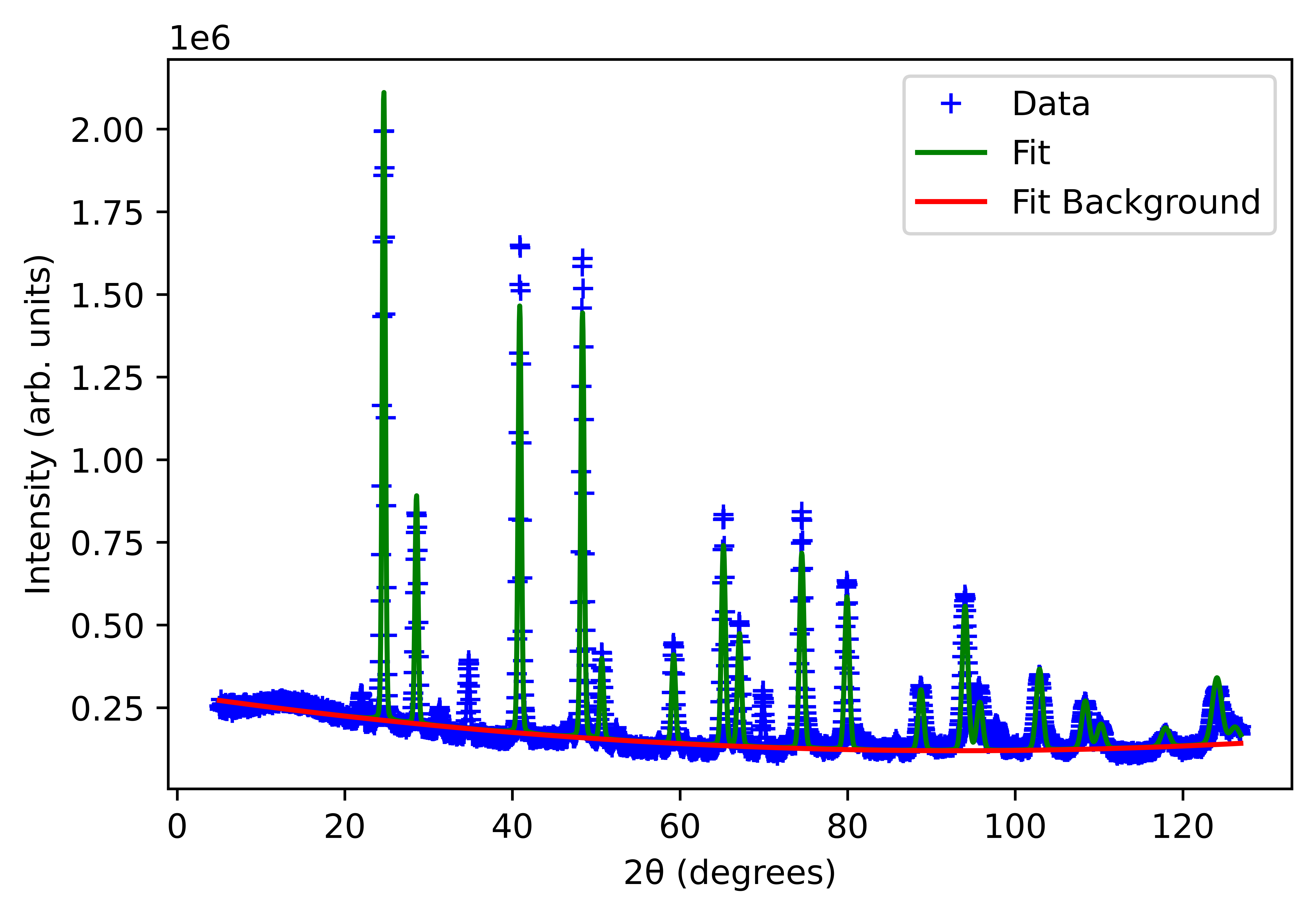}
	\caption{\label{fig:neutronRietveld} Rietveld refinement of background-subtracted neutron scattering data at 295~K, performed in GSAS II.}
\end{figure}

\begin{figure}
	\includegraphics[width=120mm]{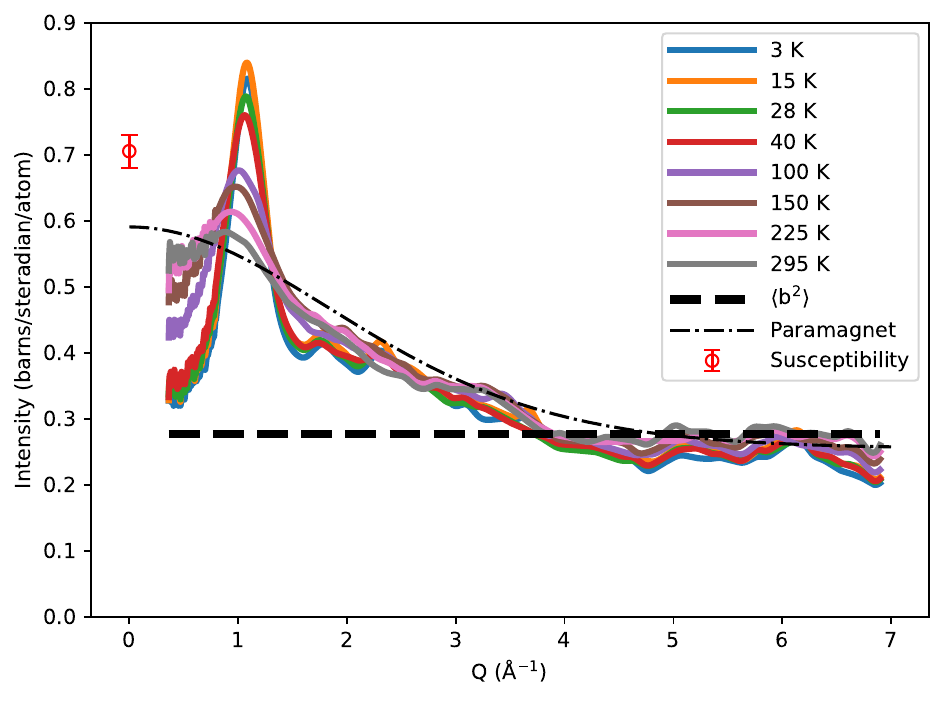}
	\caption{\label{fig:allpatterns} Diffuse scattering patterns for \znmnte\ at various temperatures, normalized to absolute units through conversion of the GSAS-II scale factors determined from Rietveld refinements. The horizontal dashed line shows the squared neutron scattering length averaged over the sample composition (which sets the expected high-$Q$ asymptote), and the dashed-dotted line is the expected scattering contribution from an ideal paramagnet. The red symbol marks the expected $Q=0$ paramagnetic scattering intensity corresponding to the effective moment size determined from the Curie-Weiss fits to the susceptibility data.}
\end{figure}

Following Ref.~\cite{paddi;jac25}, the scattered intensity was converted to a differential scattering cross section in absolute units of barns per steradian per atom by dividing the intensity by  $2 \pi^2 N C/ (4500 \lambda^3)$, where $N=8$ is the number of atoms per formula unit, $C$ is the GSAS-II refinement scale factor (refinement shown in Fig. \ref{fig:neutronRietveld}), and $\lambda$ is the neutron wavelength.
In Fig.~\ref{fig:allpatterns}, we plot the diffuse scattering patterns extracted using the algorithm described in the main article.
To check that the conversion to absolute units was successful, we compared the high-$Q$ trend of the scattering patterns to $\langle b^2 \rangle=0.277$ barns, the squared nuclear scattering length averaged over the sample composition (horizontal dashed line in the figure) The agreement is within 8\% of the average measured intensity beyond 5~\AA$^{-1}$, indicating a successful normalization to absolute units. Additionally, we show the expected scattering contribution from an ideal paramagnet (i.e., a completely random spin configuration) as the dashed-dotted curve in Fig.~\ref{fig:allpatterns}, shifted vertically upward by the averaged intensity value beyond 5~\AA$^{-1}$. We use $S=2.18$ as the effective spin, which was determined from the Spinvert fits described in the main article. The matching intensity scale and $Q$-dependence between the data and the calculated paramagnetic scattering confirms the magnetic nature of the diffuse scattering. The $Q=0$ value determined from the Curie-Weiss fits to the susceptibility is also reasonably close to the observed intensity scale. We note that the patterns and calculated paramagnetic scattering in Fig.~\ref{fig:allpatterns} are normalized per atom, whereas Fig.~7(c) in the main article shows the differential cross section normalized per Mn atom, causing it to be a factor of 4 greater than seen here.

\section{Checking the magnitude of the local magnetic order parameter}
\begin{figure}
	\includegraphics[width=80mm]{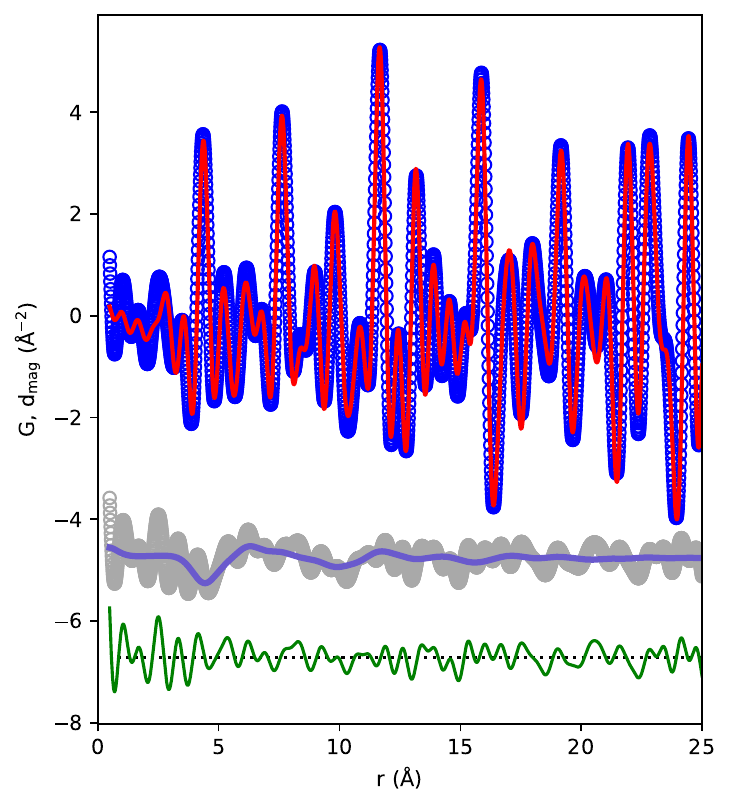}
	\caption{\label{fig:totpdf} Total PDF data for \znmnte\ at 3~K generated from the diffraction pattern collected with 1.12~\AA\ neutrons. The upper set of curves contains the experimental data (blue circles) and best-fit total PDF, comprised of the calculated atomic and magnetic PDF signals added together. Offset vertically below is the experimental mPDF signal in gray, obtained as the difference between the observed total PDF and the best-fit calculated atomic PDF, and the best-fit mPDF in purple. The overall fit residual is shown in green, further offset below.}
\end{figure}

We collected one diffraction pattern for \znmnte\ at 3~K using the short wavelength option ($\lambda$~=~1.12~\AA) at HB2A, yielding a larger maximum $Q$ value of 9.8~\AA$^{-1}$ than was available for the measurements using 1.54~\AA\ neutrons. After subtracting the instrumental background, we Fourier transformed the diffraction pattern to yield the total PDF, which includes both the nuclear (i.e. atomic) and magnetic PDF components. This total PDF pattern is shown by the blue symbols in Fig.~\ref{fig:totpdf}.
The limited $Q$ range results in a very low-resolution atomic PDF which would not be suitable for investigating subtle details of the local atomic structure, but it is sufficient for our purpose of fitting a model of the known atomic structure and subtracting out the fitted atomic PDF component to isolate the mPDF component. The difference between the total PDF and the fitted atomic PDF is shown by the gray curve, offset below the overlaid blue and red curves. This gray curve represents the mPDF, together with any noise and systematic misfits from the atomic PDF model. Fitting a model of short-range-correlated type-III antiferromagnetism to the gray curve produces the purple calculated mPDF overlaid on the gray experimental mPDF signal. The model captures the overall shape of the experimental curve quite well, indicating that we are performing a meaningful fit to the mPDF data. The high-frequency wiggles in the gray curve cannot be magnetic in origin, since the magnetic form factor restricts the magnetic scattering to relatively low $Q$ and therefore low frequency in real space. The total PDF fit, consisting of the calculated best-fit nuclear PDF and mPDF, is shown by the red curve overlaid on the blue total PDF signal in Fig.~\ref{fig:totpdf}. The overall fit residual is shown by the green curve offset below.

The local magnetic order parameter (LMOP), i.e. the magnitude of the correlated component of nearest-neighbor magnetic moments, can be calculated in units of $\mu_{\mathrm{B}}$ as
\begin{equation}
\mathrm{LMOP} = g J \sqrt{\frac{C_m \langle b \rangle ^2}{C_a n_m}}\exp\left({\frac{-r_{NN}}{2\xi}}\right),
\end{equation}
where $g$ is the Land\`e g-factor (2 in this case), $J$ is the magnitude of the angular momentum vectors (i.e., spins) used in the calculation of the mPDF, $C_m$ is the best-fit mPDF scale factor, $C_a$ is the best-fit atomic PDF scale factor, $\langle b \rangle$ is the average nuclear scattering length of the material, $n_m$ is the fraction of atoms in the material carrying a magnetic moment (0.25 in this case), $r_{NN}$ is the distance separating nearest-neighbor spins, and $\xi$ is the magnetic correlation length. Plugging the appropriate values into this equation yields $m = 3.2(3)~\mu_{\mathrm{B}}$ at 3~K, consistent with the 3.1(1)~\muB\ determined from fits to the mPDF data generated from the longer-wavelength scattering patterns.

%